\documentclass{PoS}

\usepackage{amssymb}
\usepackage{amsmath}
\usepackage{graphicx}
\usepackage{bm}

\def\lsim{\mathrel{\mathpalette\@versim<}}
\def\gsim{\mathrel{\mathpalette\@versim>}}
\def\@versim#1#2{\vcenter{\offinterlineskip
\ialign{$\m@th#1\hfil##\hfil$\crcr#2\crcr\sim\crcr } }}
\catcode`\@=12

\newcommand{\Nc}{N_{\rm c}}

\newcommand{\p}{\partial}

\newcommand{\al}[1]{\begin{align}#1\end{align}}
\newcommand{\nn}{\nonumber\\}

\newcommand{\del}{\partial}

\newcommand{\df}{\text{d}}

\newcommand{\bs}[1]{\boldsymbol}

\newcommand{\pmat}[1]{\begin{pmatrix}#1\end{pmatrix}}

\newcommand{\fn}[1]{\!\left(#1\right)}

\title{Gauge hierarchy problem and scalegenesis}

\ShortTitle{Gauge hierarchy problem and scalegenesis}

\author{\speaker{Masatoshi Yamada}
\\
        Institut f\"ur Theoretische Physik, Universit\"at Heidelberg,
Philosophenweg 16, 69120 Heidelberg, Germany\\
        E-mail: \email{m.yamada@thphys.uni-heidelberg.de}}

\abstract{
We review the gauge hierarchy problem in the standard model.
We discuss the meaning of the quadratic divergence in terms of the Wilsonian renormalization group.
Classical scale symmetry, which prohibits dimensionful parameters in the bare action, could play a key role for the understanding of the origin of the electroweak scale.
We discuss the scale-generation mechanism, i.e. scalegenesis in scale invariant theories.
In this paper, we introduce a scale invariant extension of the SM based on a strongly interacting scalar-gauge theory.
It is discussed that asymptotically safe quantum gravity provides a hint about solutions to the gauge hierarchy problem.
}

\FullConference{
Corfu Summer Institute 2019 "School and Workshops on Elementary Particle Physics and Gravity" (CORFU2019)\\
		31 August - 25 September 2019\\
		Corf\`u, Greece}

\begin{document}

\section{Introduction}
The gauge hierarchy problem~\cite{Gildener:1976ai,Weinberg:1978ym} has been a long standing problem in the standard model (SM).
The discovery of the Higgs boson with a mass of $125$\,GeV~\cite{Aad:2012tfa,Chatrchyan:2012xdj}, together with the lack of signals of new physics, implies that the SM could be valid up to a high scale.
In particular, the renormalization group (RG) analysis indicates that there is no obstacle to prevent the validity of the SM up to the Planck scale.
This situation makes the gauge hierarchy problem more mysterious.

One of aspects of the gauge hierarchy problem relies on the interpretation of the quadratic divergence.
Quantum corrections to the Higgs mass parameter contain both the logarithmic and quadratic divergences when employing a naive momentum cutoff or the Pauli-Villars regularization.
The former divergence is proportional to the bare mass parameter, while the latter one is independent from it.
The appearance of the quadratic divergence, however, strongly depends on a regularization scheme.
Indeed, the dimensional regularization does not yield a corresponding object to the quadratic divergence.
In the sense that physics should not depend on a choice of the regularization scheme, the quadratic divergence may be spurious.

The remaining gauge hierarchy problem after removing the quadratic divergence arises from the fact that the rapid energy scaling of the Higgs mass parameter: The dimensionless mass parameter behaves as $\bar m_H^2(k) \sim k^{-2}$, where $k$ is an energy scale.
This implies that the ratio between the Higgs mass parameter and the Planck scale has to be $m_H^2(M_\text{pl})/M_\text{pl}^2\sim 10^{-36}$ in order to realize $m_H^2(v_h)/v_h^2\sim 1$ at the electroweak scale.
Why is the Higgs mass at the Planck scale so much smaller than the Planck scale?
This question originates from the canonical dimension $2$ of the scalar mass parameter.
Although quantum effects induce an anomalous dimension $\gamma_m$ which deviates the energy scaling from the canonical one such that $2-\gamma_m$, it is negligibly small in the SM.
A possible solution to this problem is the imposition of classical scale symmetry which prohibits dimensionful parameters in the bare action~\cite{Wetterich:1983bi,Bardeen:1995kv}.
Then, there is no corresponding scale to the electroweak one, so that we need a scale-generation mechanism, i.e. scalegenesis.
One of possible ways is the Coleman-Weinberg mechanism~\cite{Coleman:1973jx} in which scale symmetry is broken by the scale anomaly.
The other scenario for scalegenesis relies on the strong dynamics like quantum chromodynamics.
In both cases, a degree of freedom of a dimensionless coupling changes to that of a dimensionful parameter.
This is the so-called dimensional transmutation.

The scale invariant SM, however, cannot realize electroweak scalegenesis, so that a scale invariant extension of the SM is required.
The simplest extension is an introduction of a scalar field coupled to the Higgs field via the Higgs-portal coupling.
If the dynamics in the new (hidden) sector generates a TeV scale, the electroweak symmetry breaking is triggered through the Higgs-portal coupling. 
It has been suggested a lot of possible scale invariant extensions as a hidden sector, together with other issues in the SM such as dark matter, neutrino masses and Baryogenesis.

However, classical scale invariance is nothing but a strong assumption.
The clarification of the origin of classical scale invariance requires discussing the high energy physics including quantum gravity beyond the Planck scale.
In this paper, we focus on the asymptotic safety scenario of quantum gravity.
Asymptotically safe quantum gravity is formulated as a non-perturbative quantum field theory.
An advantage of this scenario is that one can evaluate quantum gravity effects within the quantum field theory realm.
With helps of the Wilsonian RG, we can calculate a large anomalous dimensions for a scalar mass parameter, $\gamma_m$ induced by graviton fluctuations.
If $\gamma_m$ becomes larger than the canonical dimension $2$, the energy scaling behavior of the scalar mass parameter is drastically changed.
In such a scenario, we could have hints towards the gauge hierarchy problem.

This paper is organized as follows: In Section~\ref{Sec: Scalar mass and quadratic divergence in a scalar theory}, we discuss renormalization using a simple scalar field theory.
In particular, we explain the meaning of the quadratic divergence in viewpoint from the Wilson RG.
Then, we discuss the gauge hierarchy problem and the idea of classical scale invariance in Section~\ref{Sec: Gauge hierarchy problem in the SM and classical scale invariance}.
Section~\ref{Section: Scalegenesis} is devoted to introducing a mechanism for scalegenesis and to showing an example of an extension of the SM.
In Section~\ref{Sec: Asymptotically safe quantum gravity and gauge hierarchy problem}, we discuss impacts of graviton fluctuations in asymptotically safe quantum gravity on the scalar mass parameter, especially the anomalous dimension $\gamma_m$ induced by graviton fluctuations, and potential scenarios as a solution to the gauge hierarchy problem.

\section{Scalar mass and quadratic divergence in scalar theory}
\label{Sec: Scalar mass and quadratic divergence in a scalar theory}
In this section we discuss renormalization for a simple real singlet-scalar field theory.
We start by reviewing the mass-independent renormalization (MIR) in the standard perturbation theory.
It is shown in Section~\ref{Sec: Mass-independent renormalization} that the scalar mass parameter is separated into the logarithmic and quadratic divergent parts.
One can deal ``multicatively" with the logarithmic divergence in renormalization, while the quadratic one is ``additively" renormalized. 
In Section~\ref{Sec: Wilsonian renormalization group} we discuss the Wilson RG procedure for the same scalar theory.
In particular, we highlight that the quadratic divergence in terms of the Wilsonian RG specifies an ``absolute" position of the phase boundary between the broken and symmetric phases in the theory space for a chosen renormalization scheme.  
The independence of physical quantities from the choice of the renormalization scheme requires that they are defined as a deviation (relative distance) from the phase boundary.
This can be actually seen by analyzing the RG flow around a fixed point.
Then, the quadratic divergence becomes invisible by rotating the coordinate of the theory space around the fixed point.

\subsection{Mass-independent renormalization}
\label{Sec: Mass-independent renormalization}
We demonstrate the MIR scheme using a real singlet-scalar theory whose Lagrangian reads
\al{
{\mathcal L}= \frac{1}{2}(\p_\mu \phi_0)^2 -\frac{1}{2} m_0^2\phi_0^2 -\frac{1}{4!}\lambda_0 \phi_0^4\,,
}
where the subscript ``0" denotes a bare quantity.
A key treatment of the MIR is that we split the bare mass into two parts, i.e. $m_0^2=\widehat m_0^2+\delta m_0^2$, and define renormalized quantities such that
\al{
&\phi=Z_\phi ^{1/2}\phi_0\,,& 
&m^2=Z_m^{-1}\widehat m_0^2\,,& 
&\delta m^2=Z_\phi \delta m_0^2\,,&
&\lambda=Z_\lambda^{-1}\lambda_0\,.
\label{Eq: definition of renormalized quantities}
}
We should note here that the renormalized mass $m^2$ is proportional to $\widehat m_0^2$ but not to $\delta m_0^2$.
For this setup the Lagrangian with counter terms in terms of the renormalized quantities \eqref{Eq: definition of renormalized quantities} is given by
\al{
{\mathcal L}= \frac{1}{2}(\p_\mu \phi)^2 -\frac{1}{2} m^2\phi^2 -\frac{1}{4!}\lambda\phi^4 + {\mathcal L}_{\rm c.t.}\,,
}
where the counter-term Lagrangian reads
\al{
{\mathcal L}_{\rm c.t.}= \frac{\delta Z_\phi}{2}(\p_\mu \phi)^2 -\frac{\delta Z_m}{2} m^2\phi^2-\frac{1}{2} \delta m^2\phi^2 -\frac{\delta Z_\lambda}{4!}\lambda\phi^4 \,,
}
with $\delta Z_\phi=Z_\phi -1$, $\delta Z_m=Z_mZ_\phi-1$ and $\delta Z_\lambda=Z_\lambda Z_\phi^2-1$.
To determine each counter term we employ the following four renormalization conditions:
\al{
&\Gamma^{(2)}(p^2=0, m^2,\lambda; \mu^2)\Big|_{m^2=0}=0\,, 
\label{R condition for quadratic mass}
\\[1ex]
&\displaystyle \frac{\p}{\p m^2}\Gamma^{(2)}(p^2=0, m^2,\lambda; \mu^2)\Big|_{m^2=\mu^2}=-1\,,
\label{R condition for mass}
\\[1ex]
&\displaystyle \frac{\p}{\p p^2}\Gamma^{(2)}(p^2, m^2,\lambda; \mu^2)\Big|_{p^2=0,\,m^2=\mu^2}=1\,,
\label{R condition for field renormalization}
\\[2ex]
&\Gamma^{(4)}(\{p_i\}=0,m^2,\lambda;\mu^2)|_{m^2=\mu^2}=-\lambda\,,
}
where $\mu$ is the renormalization scale; $\Gamma^{(2)}$ and $\Gamma^{(4)}$ stand for the two- and four-point functions of the scalar field, respectively; $p$ is an external momentum; and $\{p_i\}$ denotes a set of external momenta, e.g. $\{p_i\}=(p_1,p_2,p_3,p_4)$ for the four-point function.\footnote{
We note that one of momenta is redundant due to the momentum conservation.
}

Let us now calculate the two-point function.
At the one-loop level, one can obtain
\vspace{-3mm}
\al{
\Gamma^{(2)}(p^2)\Big|_{\mathcal O(\lambda^1)}
&=~~\vcenter{\hbox{\includegraphics[width=70mm]{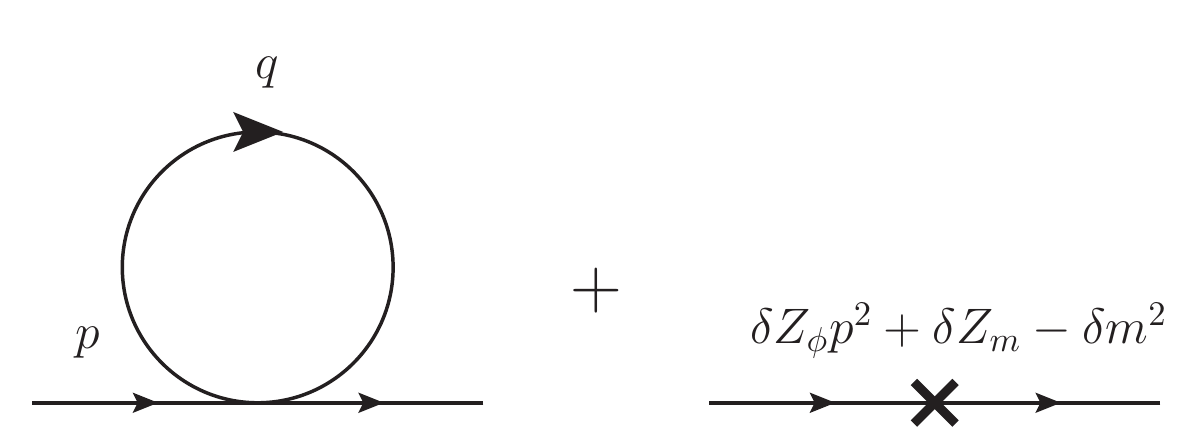}}\vskip5mm}\nn[1ex]
&=-\frac{i\lambda}{2}\int \frac{\df^4 q}{(2\pi)^4}\frac{1}{q^2-m^2}~~+~~\delta Z_\phi^{(1)} p^2 - \delta Z_m^{(1)}m^2 - (\delta m^2)^{(1)} \,.
}
Here the one-loop integration is calculated as 
\al{
-\frac{i\lambda}{2}\int& \frac{\df^4 q}{(2\pi)^4}\frac{1}{q^2-m^2}\nn
&\quad
=\begin{cases}
\displaystyle -\frac{\lambda}{32\pi^2}(2\ln2 )\Lambda^2 + \frac{\lambda}{32\pi^2}m^2\left( 1+ \ln\frac{\Lambda^2}{m^2}\right)  & \text{(Pauli-Villars reg.)} \\[3ex]
\phantom{\displaystyle -\frac{\lambda}{32\pi^2}(2\ln2 )\Lambda^2}\displaystyle + \frac{\lambda}{32\pi^2}m^2\left( 1 +\frac{1}{\epsilon} \right) & \text{(dimensional reg.)}
\end{cases}\,,
}
where $\Lambda$ is a UV cutoff in the Pauli-Villars regularization,\footnote{
To regularize the UV divergence, we employ the following replacement for the scalar propagator: 
\al{
\frac{1}{q^2-m^2}\to
\left(\frac{1}{q^2-m^2} -\frac{1}{q^2-\Lambda^2}\right) -\left( \frac{1}{q^2-(\Lambda^2+m^2)} - \frac{1}{q^2-2\Lambda^2} \right)\,.
}
}
and we defined $1/\epsilon = 2/(4-d)-\gamma_E +\ln(4\pi)$ with $\gamma_E$ the Euler gamma constant and $d$ spacetime dimension.
From the conditions \eqref{R condition for mass} and \eqref{R condition for field renormalization}, we can determine the field and mass renormalization factors,
\al{
&\delta Z_\phi^{(1)}=0 \,,&
&\delta Z_m^{(1)}=\frac{\lambda}{32\pi^2} \left( 1 +\left\{ \ln\frac{\Lambda^2}{\mu^2}~~\text{or}~~  \frac{1}{\epsilon} \right\}  \right) \,,
}
while the first condition \eqref{R condition for quadratic mass} gives
\al{
(\delta m^2)^{(1)}
=\begin{cases}
\displaystyle -\frac{\lambda}{32\pi^2}(2\ln2) \Lambda^2 & \text{(Pauli-Villars reg.)}\\[2ex]
0 & \text{(dimensional reg.)}
\end{cases}.
}
We see that $\delta m^2$ subtracts the quadratic divergence $\Lambda^2$ in the Pauli-Villars regularization.
Thus, the MIR scheme allows us to separate the bare mass into the logarithmically divergent part and the quadratically divergent part.
In the dimensional regularization the quadratic divergence is automatically subtracted without the introduction of $\delta m^2$.
Note here that $Z_\phi$ and $Z_m$ do not depend on the renormalized mass parameter $m^2$.
Indeed, $Z_\lambda$ is also independent from $m^2$.
This is why this renormalization scheme is called the ``mass-independent" renormalization.
Note also that it is guaranteed in a fermionic theory that the bare fermion mass term $\delta m_0$ is zero in any regularization scheme thanks to chiral symmetry, so that the bare mass is always proportional to the renormalized mass.
In a gauge theory, the gauge symmetry realizes the same situation.

Once one obtains the renormalization factors, one can evaluate the anomalous dimensions which characterize the energy scaling of coupling constants,  
\al{
&\beta(\lambda) = -\lambda \mu \frac{\p \ln Z_\lambda}{\p \mu}\,,&
&\gamma_m(\lambda) = -  \mu \frac{\p \ln Z_m}{\p \mu}\,.&
&\gamma_\phi(\lambda) = \frac{1}{2} \mu \frac{\p \ln Z_\phi}{\p \mu}\,.
\label{Def of anomalous dimensions}
}
From the second definition in Eq.\,\eqref{Def of anomalous dimensions} and the definition of the renormalized scalar mass, one obtains 
\al{
\mu \frac{\p}{\p \mu} {\bar m}^2&= \left( -2 + \gamma_m\right){\bar m}^2\,,
\label{perturbative mass RG equation}
}
where $\bar m^2=m^2/\mu^2$ is the dimensionless scalar mass, and the anomalous dimension at the one-loop level reads
\al{
\gamma_m=\frac{1}{16\pi^2}\lambda\,.
} 
Since $Z_\phi=1$ at the one-loop level, one has $\gamma_\phi=0$.
The RG equation for the quartic coupling constant at the one-loop level is well known as
\al{
\mu \frac{\p}{\p \mu} \lambda &= \frac{3\lambda^2}{16\pi^2}\,.
}
The solution to this equation behaves $\lambda \to0$ for $\mu\to0$, while there is a Landau pole at $\mu=\mu_0 e^{\frac{16\pi^2}{3\lambda(\mu_0)}}$ where $\lambda(\mu_0)$ is a boundary value at $\mu=\mu_0$.

\subsection{Wilsonian renormalization group}
\label{Sec: Wilsonian renormalization group}
Let us briefly introduce the Wilsonian RG in quantum field theory.
The task in quantum field theory is to solve the path integral which represents the summation of all quantum fluctuations.
We now consider to divide quantum fluctuations into high momentum modes and low momentum modes by introducing an infrared (IR) cutoff scale $k$ such that $\phi(p)=\phi_>(p)+\phi_<(p)$ where $\phi_>(p)$ and $\phi_<(p)$ are fields with higher momentum modes $|p|>k$, and lower momentum modes $|p|>k$, respectively.
Thus, we perform the path integral only for the field $\phi_>(p)$ and define an effective action which is a functions of $\phi_<(p)$.
This operation is the so-called coarse-graining.
Varying the IR cutoff scale to $k=0$, we can obtain the full effective action including all quantum fluctuations.
Such a process can be formulated as a functional differential equation.
One of formulations is given for the one-particle irreducible (1PI) effective action $\Gamma_k$~\cite{Wetterich:1992yh} so that
\al{
\p_t \Gamma_k = \frac{1}{2} \text{Tr} \left[ \left( \Gamma_k^{(2)} +R_k \right)^{-1}\cdot \p_t R_k \right]\,,
\label{Wetterich equation}
}
where $R_k(p)$ is a cutoff profile function (or simply a regulator) realizing the coarse-graining, $\Gamma_k^{(2)}$ is the full two-point function, namely $\Gamma_k^{(2)}=\frac{\delta^2}{\delta \phi^2}\Gamma_k$, and ``Tr" denotes the functional trace acting on all internal spaces in which the field is defined, e.g. momentum and flavor, etc.

The 1PI effective effective action is in general spanned by an infinite number of effective operators ${\mathcal O}_i(x)$:
\al{
\Gamma_k=\int \df^4x \sum_i^\infty\,  {g}_i\,{\mathcal O}_i(x)\,,
}
where $g_i$ is a dimensionful coupling constant.
Using the flow equation \eqref{Wetterich equation} and defining a dimensionless coupling constant $\bar g_i= k^{-(4-D_i)} g_i$, where $D_i$ is the mass dimension of $\mathcal O_i$, one obtains the RG equation for ${\bar g}_i$,
\al{
\p_t {\bar g}_i =\beta_i (\{ {\bar g}\})\,,
\label{RG equation in general}
}
where $\{ {\bar g}\}$ denotes a set of coupling constants.
The right-hand side of Eq.\,\eqref{RG equation in general} is the beta function.

One of important things in the RG is the existence of a fixed point $\bar g_{i*}$ at which all beta functions vanish, i.e. $\beta_i (\{ {\bar g}_{*}\})=0$ for all $i$.
Once a fixed point is found, one can analyze the RG flows around the fixed point.
To this end, we expand the beta functions in Eq.\,\eqref{RG equation in general} around a fixed point $g_{i*}$ and take into account only the linear term,
\al{
\p_t ( {\bar g}_i -{\bar g}_i^* )\simeq  -T_{ij}({\bar g}_j-{\bar g}_{j*})\,,
\label{linearized RG equation in a general theory}
}
where we have defined the stability matrix,
\al{
T_{ij}=-\frac{\p \beta_i}{\p {\bar g}_j}\bigg|_{g=g_*} \,.
\label{Def of stability matrix}
}
The equation \eqref{linearized RG equation in a general theory} can be solved easily:
Diagonalizing the stability matrix \eqref{Def of stability matrix} and denoting its eigenvalues by $\theta_i$, one finds
\al{
{\bar g}_i-{\bar g}_{i*}= \sum_j V_{ij}C_j \left( \frac{k}{\Lambda} \right)^{-\theta_j}\,.
\label{solution of RGE in general}
}
Here, $V_{ij}$ is a matrix diagonalizing the stability matrix \eqref{Def of stability matrix} and $C_j$ is a constant at a reference scale $\Lambda$.
When the matrix $V$ is approximated to be diagonal, $V_{ij}\approx v_i\delta_{ij}$, one has
\al{
{\bar g}_i-{\bar g}_{i*}= v_i\, C_i \left( \frac{k}{\Lambda} \right)^{-\theta_i}\,.
\label{solution of RGE in approximated}
}
One can see from Eq.\,\eqref{solution of RGE in approximated} that $\theta_i$ characterizes the energy scaling of coupling constant ${\bar g}_i$ and is called as the ``critical exponent".
In particular, for a found fixed point, the sign of the critical exponents is important.
There are the following cases:
\al{
&\theta_i >0 \qquad \text{relevant}\,,\nn
&\theta_i =0  \qquad \text{marginal}\,,\\
&\theta_i <0  \qquad \text{irrelevant}\,.\nonumber
}
Towards IR regimes, the RG flow of a relevant coupling ($\theta_i>0$) goes away from the fixed point, while that of an irrelevant one ($\theta_i<0$) shrinks to the fixed point.
This fact leads to the notion of the renomalizability of a theory.
Since irrelevant couplings, starting at an arbitrary value in a UV scale, converge to the fixed point value in the low energy region, they are predictable, i.e. $\bar g_{i}=\bar g_{i*}$.
If a theory has a finite number of relevant couplings, the theory is renormalizable.

A typical structure of the beta functions becomes the following form:
\al{
\p_t \bar g_i = -(4-D_i)\bar g_i + f_i(\{ \bar g \})\,,
\label{typical structure of the beta functions}
}
where $D_i$ is the mass-dimension of a corresponding operator to $g_i$.
The first term on the right-hand side is the canonical scaling term reflecting the mass-dimensionality of the coupling $\bar g_i$, and the second one involves quantum effects.
Let us here assume that the system described by Eq.\,\eqref{typical structure of the beta functions} has a fixed point $\bar g_{i*}$.
In this case, the stability matrix is given by
\al{
T_{ij}= (4-D_i)\delta_{ij} + \frac{\p f_i}{\p \bar g_j}\bigg|_{g=g_*}\,,
\label{Eq: stability matrix 2}
}
from which the critical exponent is obtained as $\theta_i=\text{eig}_i(T)$.

In particular, for the Gaussian fixed point $\bar g_{i*}=0$ or a small one $\bar g_{i*}\ll1$, the quantum effects (second term on the right-hand side in Eq.\,\eqref{Eq: stability matrix 2}) are negligible, so that the critical exponent is identical with the canonical scaling, i.e. $\theta_i =4-D_i$.
This is consistent with the standard definition of the renormalizability in the perturbation theory.
One can see the critical exponents at the Gaussian fixed point from a naive dimensional counting of coupling constants, whereas at a non-vanishing fixed point, in general, the second term on the right-hand side in Eq.\,\eqref{Eq: stability matrix 2} is finite, so that the critical exponents deviate from the canonical scalings of coupling constants, i.e.  the second term is the anomalous dimension induced by quantum effects.

We now analyze a scalar theory using the flow equation \eqref{Wetterich equation}.
Quantum effects generate an infinite number of effective operators which respect with symmetries the bare action has.
We assume that the scalar theory has the $Z_2$ symmetry which admits only even powers of $\phi$ in the effective action.
For the present purpose we take only the following three terms into account and truncate higher operators:
\al{
\Gamma_k[\phi] \simeq \int \df^4 x \left[ \frac{Z_{\phi}^{-1}}{2}(\p_\mu \phi)^2  + \frac{1}{2}m^2  \phi^2 + \frac{1}{4!}\lambda \phi^4\right]\,,
}
where we consider the Euclidean spacetime.
Using the flow equation \eqref{Wetterich equation} one obtains
\al{
\p_t \bar m^2 &= -2(1-\gamma_\phi ){\bar m}^2 - \frac{\bar \lambda}{32\pi^2} \ell_1^4({\bar m}^2)\,,
\label{FRGE for mass}
\\[2ex]
\p_t \bar \lambda &= 4\gamma_\phi {\bar \lambda} +  \frac{3\bar \lambda^2}{16\pi^2} \ell_2^4({\bar m}^2)\,,
\label{FRGE for quartic coupling}
}
where a bar denotes a dimensionless quantity, the first terms on the right-hand side of Eqs.\,\eqref{FRGE for mass} and \eqref{FRGE for quartic coupling} are canonical scaling terms which reflect the canonical dimensions of the scalar mass term and the quartic coupling, and the second terms are loop effects.
See e.g.~Ref.\,\cite{Berges:2000ew} for the derivation of these RG equations.
Here, $\gamma_\phi= \p_t \ln Z_{\phi}/2$ is the anomalous dimension from the field renormalization factor and we define the threshold functions in four dimensional spacetime,
\al{
&\ell_0^4(\tilde w)= k^{-2}\int_0^{\Lambda^2} \df x\,\frac{x}{2} \frac{\p_t R_k(x)}{x + R_k(x)+ \tilde w}\,,&
&\ell_n^4(\tilde w)= (-1)^n \frac{1}{n!}\frac{\p^n}{\p \tilde w} \ell_0^4(\tilde w)\,.
\label{Eq: threshold function}
}
For the Litim-type cutoff function~\cite{Litim:2001up},
\al{
R_k(p^2)=(a\, k^2-p^2)\theta(a\, k^2-p^2)\,,
\label{Eq: Litim cutoff}
}
we can perform the momentum integral in the threshold function \eqref{Eq: threshold function} and then obtain
\al{
\ell_n^4({\bar m}^2) = \frac{a^{-n+2}}{(1+ a^{-1}\bar m^2)^{n+1}}\,,
}
where $a$ is a (dimensionless) positive and finite constant and describes a class of regularization schemes.
Note here that $\ell_2^4(0)$ is independent from the parameter $a$.

Let us solve the RG equations \eqref{FRGE for mass} and \eqref{FRGE for quartic coupling}.
For simplicity, we assume that $a^{-1}\bar m^2 \ll1$, i.e. we can expand the threshold functions as 
\al{
&\ell_1^4({\bar m}^2)\approx \ell_1^4(0)- 2\ell_{2}^4(0){\bar m}^2 \,,&
&\ell_2^4({\bar m}^2)\approx \ell_2^4(0)\,.
}
We neglect the anomalous dimension of the field renormalization factor, $\gamma_\phi=0$ (or equivalently $Z_{\phi}^{-1}=1$).\footnote{
This approximation is called as the local potential approximation.
}
Indeed, for scalar theories in the symmetric phase, the field renormalization factor does not deviate from one.
Then, the RG equation for the scalar mass is simplified to be
\al{
\p_t \bar m^2 &= -2{\bar m}^2 - \frac{a\bar \lambda}{32\pi^2}  + \frac{\bar \lambda}{16\pi^2} {\bar m}^2\,,
\label{Wilson RG for mass}
}
where we have used $\ell_1^4(0)=a$ and $\ell_2^4(0) =1$ in the Litim-type cutoff \eqref{Eq: Litim cutoff}.
Compare this to Eq.\,\eqref{perturbative mass RG equation}.
The last term on the right-hand side of Eq.\,\eqref{Wilson RG for mass} can be identified to $\gamma_m$, while the second term is missing in Eq.\,\eqref{perturbative mass RG equation}.
This term is actually the contribution from the quadratic divergence: By solving Eq.\,\eqref{Wilson RG for mass} without the last term of Eq.\,\eqref{Wilson RG for mass}, one finds, for the dimensionful scalar mass,
\al{
m^2(k\to 0)= -\frac{a\bar\lambda}{64\pi^2}\Lambda^2 \,.
}
The Wilsonian RG does not respect to scale symmetry due to the introduction of the cutoff scale $k$, so that in general, the quadratic divergence appears in the beta function of the scalar mass.
The quadratic divergent part explicitly depends on the parameter $a$.

\begin{figure}[t]
\centering
\includegraphics[width=7.5cm]{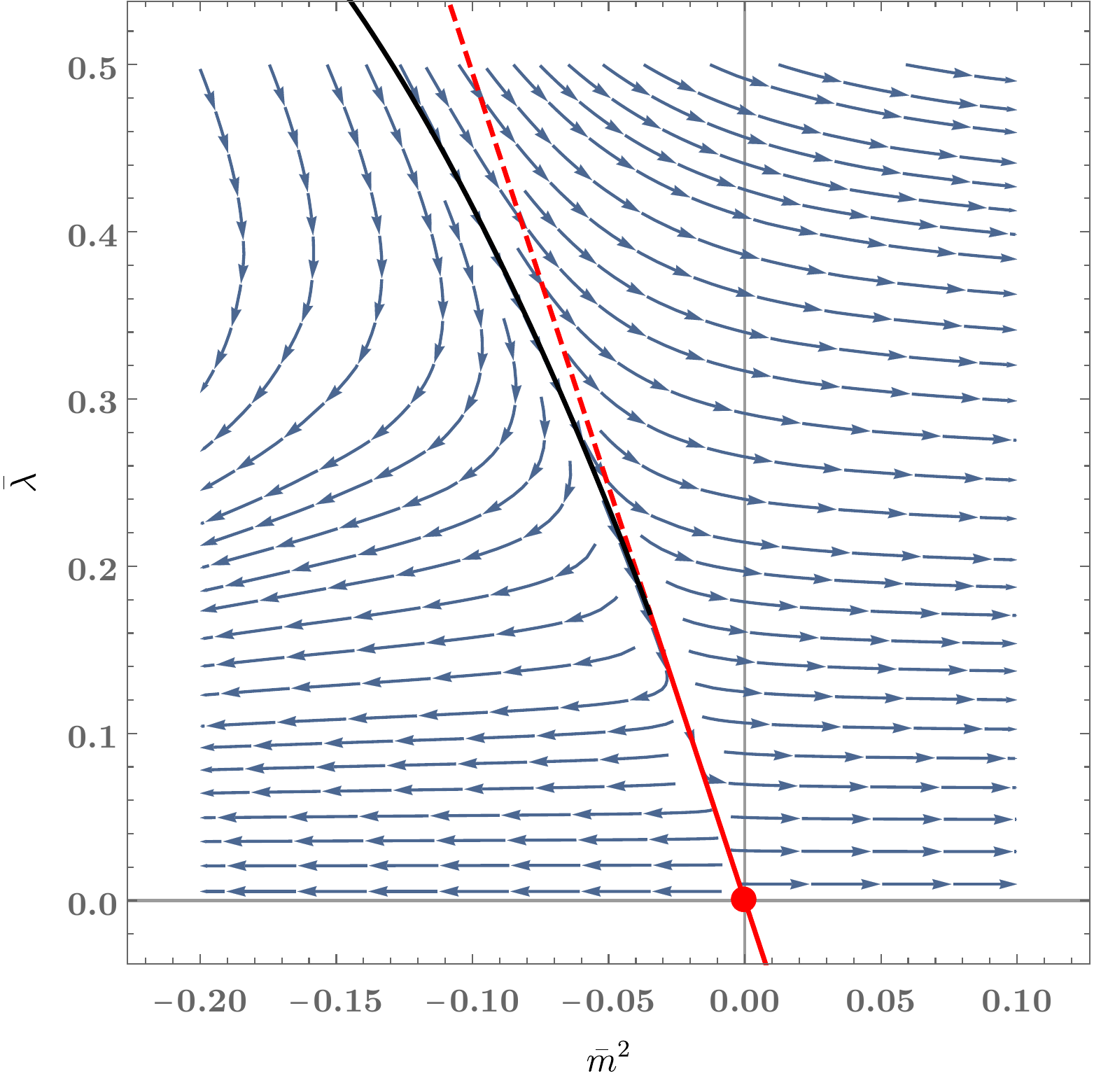}
\includegraphics[width=7.5cm]{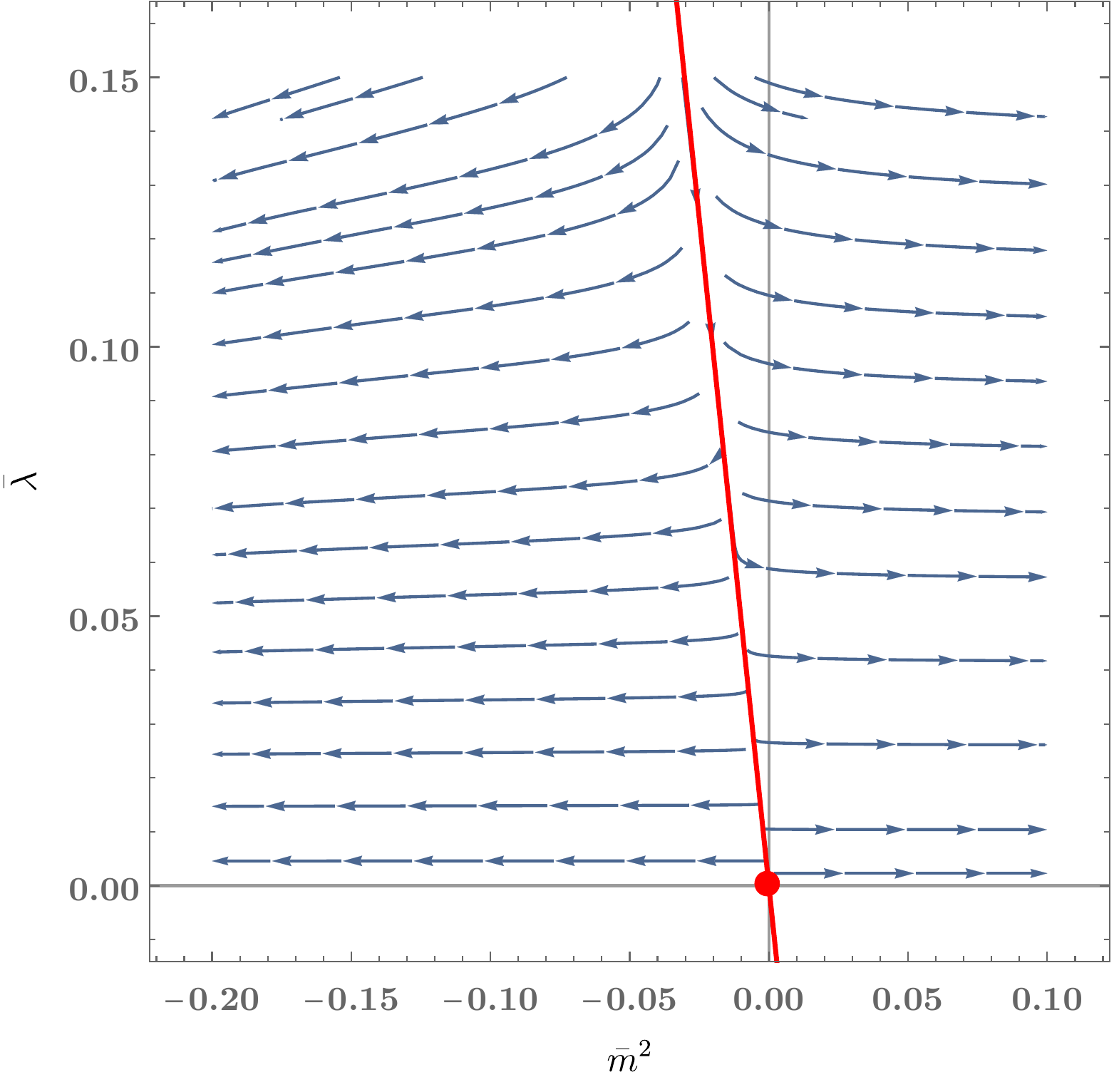}
\caption{RG flows in the $\bar m^2$-$\bar \lambda$ plane. Arrows indicate flows from UV to IR. The red point is the Gaussian fixed point ($\bar m^2_*=0$, $\bar\lambda_*=0$).
The red and black solid lines denote the phase boundaries around and apart from the Gaussian fixed point, respectively.
}
\label{phase boundary lines} 
\end{figure}

What is the meaning of the quadratic divergence in the Wilsonian RG?
To answer to this question, let us analyze the RG flow around a fixed point in the scalar system.
In a scalar theory in four dimensional spacetime, one could find, for a positive quartic coupling constant,\footnote{
If we extend the quartic coupling constant to negative values, there exists a non-trivial fixed point~\cite{Gawedzki:1985cf}.
}
only the Gaussian (trivial) fixed point, $\bar m^2_*=\bar \lambda_*=\cdots=0$ at which the linearized RG equation \eqref{linearized RG equation in a general theory} reads 
\al{
\p_t \pmat{
\bar m^2\\[1ex]
\bar \lambda
}
 \simeq  \left.\pmat{
\frac{\p \beta_m}{\p \bar m^2} & \frac{\p \beta_m}{\p \bar \lambda} \\[2ex]
\frac{\p \beta_\lambda}{\p \bar m^2} & \frac{\p \beta_\lambda}{\p \bar \lambda} 
}\right|_{\substack{\bar m_*^2=0 \\[0.5ex] \bar \lambda_*=0}}
 \pmat{
\bar m^2 \\[1ex]
\bar \lambda
}=
\pmat{
-2  & & -\frac{a}{32\pi^2}\\[2ex]
0 & &  0
}
\pmat{
\bar m^2 \\[1ex]
\bar \lambda 
}\,.
\label{linearized equation in scalar theory}
}
Note that the term $-\frac{a}{32\pi^2}$ comes from the quadratically divergent part (the second term on the right-hand side) in Eq.\,\eqref{Wilson RG for mass}.
Diagonalizing the stability matrix by the rotation matrix,
\al{
V=\pmat{
1 && -\frac{a}{64\pi^2} \\[1ex]
0 && 1
}\,,
\label{rotation matrix in an example}
} 
we find one of the linearized RG equations,
\al{
\p_t \left(\bar m^2 +\frac{a}{64\pi^2}\bar \lambda \right)&=-2 \left( \bar m^2+ \frac{a}{64\pi^2} \bar\lambda \right)\,,
\label{linearized RG diff equation}
}
and its solution,
\al{
\bar m^2(k) + \frac{a}{64\pi^2}\bar \lambda(k)= e^{2t}\left(  \bar m_0^2 + \frac{a}{64\pi^2} \bar\lambda_0 \right)\,,
\label{linearized RG equation for an example scalar theory in Wilson RG}
}
where $\bar m_0^2=\bar m^2(k=\Lambda)=m_0^2/\Lambda^2$ and $\bar \lambda_0=\bar \lambda(k=\Lambda)=\lambda_0$.
Note that $k=\Lambda$ corresponds to $t=0$.
We see that for the choice at the UV scale,
\al{
m_0^2 = -\frac{a\bar\lambda_0}{64\pi^2} \Lambda^2\,,
\label{quadratic divergence in linearized equation}
}
the RG flow \eqref{linearized RG equation for an example scalar theory in Wilson RG} gives the relation
\al{
\bar m^2(k)  =-\frac{a}{64\pi^2}\bar \lambda(k)=: \bar m_c^2(\bar\lambda)\,.
\label{linear relation between m2 and lambda}
}
Since the RG flow of the quartic coupling constant goes to zero in the IR limit $k\to 0$, one can see $\bar m^2(k\to 0)\to0$ from this equation \eqref{linear relation between m2 and lambda} .
Hence, the choice \eqref{quadratic divergence in linearized equation} is a condition to obtain the massless theory at the IR limit and the relation \eqref{linear relation between m2 and lambda} describes the RG trajectory towards the massless theory.
It is clear from Eq.\,\eqref{linearized RG equation for an example scalar theory in Wilson RG} that for a choice $m_0^2> \bar m_c^2(\bar\lambda_0)$, the scalar mass takes a positive value in the IR limit, namely the theory is in the symmetric phase in the IR regime.
In the opposite inequality case ($m_0^2< \bar m_c^2(\bar\lambda_0)$), the broken phase is observed in the IR regime. 
To summarize, the relation \eqref{linear relation between m2 and lambda} represents the phase boundary (critical line) between the symmetric and broken phases for the $Z_2$ symmetry~\cite{Wetterich:1983bi,Aoki:2012xs}.
In particular, the theory on the phase boundary is massless, i.e. {\it critical}. 
In Fig.\,\ref{phase boundary lines}, we plot the RG flows in the $\bar m^2$-$\bar\lambda$ plane.
The red line in Fig.\,\ref{phase boundary lines} is the phase boundary around the Gaussian fixed point described by Eq.\,\eqref{linear relation between m2 and lambda}.

However, the slope of this phase boundary strongly depends on the regularization scheme.
Indeed, the choice of the regularization scheme corresponds to the choice of the coordinate in the theory space~\cite{Sumi:2000xp}. 
A choice of the coordinate is connected to other choices by rotations of the coordinate.
Therefore, one can rotate arbitrary coordinates such that the quadratic divergence is subtracted. 
The matrix corresponding to such a rotation is actually given by Eq.\,\eqref{rotation matrix in an example} in the present analysis.
In this sense, the dimensional regularization is a scheme so that the rotation matrix becomes the identity matrix.

\begin{figure}[t]
\centering
\includegraphics[width=7.5cm]{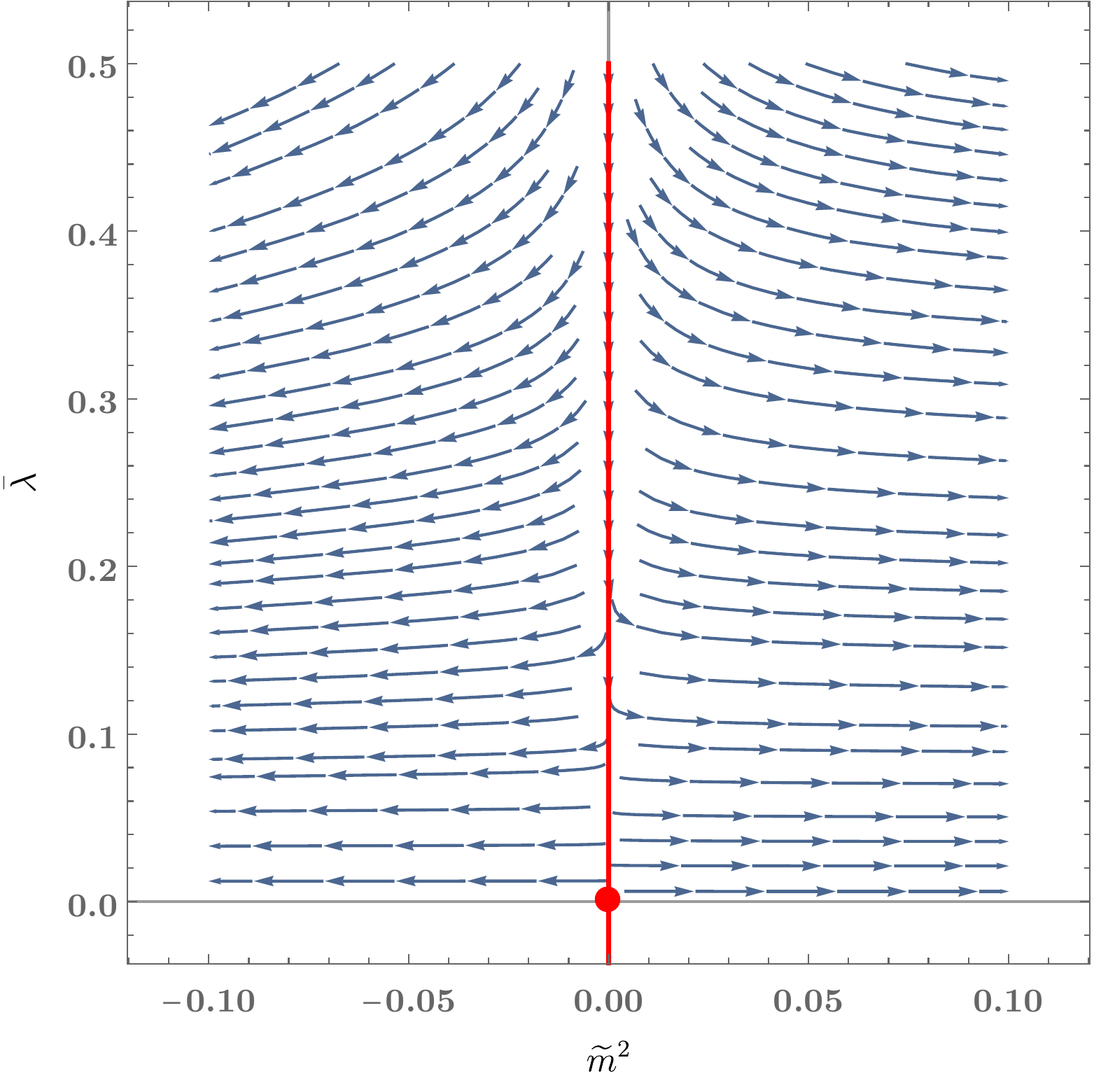}
\caption{RG flows in the $\widetilde m^2$-$\bar \lambda$ plane, where $\widetilde m^2$ is defined in Eq.\,\eqref{scalar mass in new basis}.
}
\label{phase boundary lines in new basis} 
\end{figure}

The quantities after the rotation correspond to the relative distance between the RG flow and the phase boundary (or the deviation from the phase boundary) are universal, i.e. independent from the regularization scheme.
From Eq.\,\eqref{linearized RG equation for an example scalar theory in Wilson RG} one can define
\al{
&\widehat m_0^2= m_0^2 + \frac{a\bar\lambda_0}{64\pi^2}\Lambda^2\,,&
&\widetilde m^2  \equiv \bar m^2  + \frac{a}{64\pi^2}\bar \lambda\,.
\label{scalar mass in new basis}
}
This is the scalar mass in the new basis for the theory space.
The RG equations for the scalar mass and the quartic coupling constant in the new basis are given repectively by 
\al{
&\p_t \widetilde m^2=-2\widetilde m^2\,,&
&\p_t \bar\lambda=0\,.
}
The RG flows in the new basis is represented in Fig.\,\ref{phase boundary lines in new basis}.
The one-loop effects in these equations are missing.
To take into account them, we have to evaluate non-linear terms in Eq.\,\eqref{linearized equation in scalar theory}.
For a large value of the quartic coupling constant the phase boundary is bended as one can see from the black solid line in Fig.\,\ref{phase boundary lines}.

\section{Gauge hierarchy problem in the SM and classical scale invariance}
\label{Sec: Gauge hierarchy problem in the SM and classical scale invariance}
So far, we have discussed renormalization in a simple scalar theory for both the standard perturbative renormalization and the Wilsonian RG.
In particular, we have argued the meaning of the quadratic divergence. 
In the SM with the momentum regularization, one finds, at the one-loop level,\footnote{The two-loop contribution has been computed, e.g. in Ref.\,\cite{Hamada:2012bp}.}
\al{
\delta m^2= \frac{1}{16\pi^2}\left( -6\lambda + 6y_t^2 -\frac{9}{4}g_2^2 - \frac{3}{4}g_1^2 \right) \Lambda^2\,,
\label{Eq: quadratic divergence of SM}
}
where $\lambda$ is the quartic coupling constant of the Higgs doublet-field; $y_t$ is the top-quark Yukawa coupling constant; and $g_2$ and $g_1$ are the SU(2)$_L$ and U(1)$_Y$ gauge coupling constants, respectively.
If the SM is valid up to the Planck scale, the quadratic divergence has to be precisely canceled with the bare mass $\delta m_0^2$ which does not depend on the renormalized scalar mass $m^2$.
This is the so-called gauge hierarchy problem or fine-tuning problem~\cite{Gildener:1976ai,Weinberg:1978ym}.
This problem has motivated us to consider supersymmetric extensions of the SM, in which the quadratic divergences from a particle and its superpartner cancel each other out; see e.g.~Ref.\,\cite{Martin:1997ns}.
An other idea for a solution to the gauge hierarchy problem is the Veltman condition~\cite{Veltman:1980mj} which states that the coefficient of the quadratic divergence (the combination of coupling constants in Eq.\,\eqref{Eq: quadratic divergence of SM}) should vanish at a UV scale at which the bare theory is defined.

Is the quadratic divergence, however, physically meaningful?
As discussed in the previous section, the quadratic divergence is always subtracted by the counter term $\delta m^2$.
In view point of the Wilsonian RG, the quadratic divergence specifies the position of the phase boundary and strongly depends on the choice of the coordinate (regularization scheme) in the theory space.
Rotating the coordinate around the Gaussian fixed point, the quadratic divergence is invisible in the new basis of the coordinate.
If high energy theories such as string theory determine the coordinate of the theory space, one has to seriously discuss the gauge hierarchy problem why the quadratic divergence is precisely canceled out with the bare Higgs mass.
Nevertheless, as far as one discusses the dynamics of particles within low energy effective theories, one does not have to specify the scheme-dependent coordinate of the theory space, so that the quadratic divergence may be spurious.

Once we accept the idea that the quadratic divergence is irrelevant for the low energy dynamics, we can consider only the logarithmic divergence which contributes to the running of the scalar mass.
Is the gauge hierarchy problem resolved if the quadratic divergence is subtracted?
Its answer is No.
To see the remaining problem, let us study the RG equation given in Eq.\,\eqref{perturbative mass RG equation} or Eq.\,\eqref{Wilson RG for mass} after the quadratic divergence is subtracted.
Assuming that the running effects of the anomalous dimension $\gamma_m$ is negligible, the running (dimensionless) scalar mass parameter is given by
\al{
\bar m^2(\mu) = {\bar m}_0^2 \left( \frac{\mu}{\Lambda} \right)^{-2+\gamma_m}\,,
\label{running scalar mass}
}
where $\bar m_0^2$ is the boundary mass at the UV scale, i.e. ${\bar m}_0^2=\bar m^2(\mu=\Lambda)=\widehat m_0^2/\Lambda^2$.
In the SM, the anomalous dimension at one-loop level is computed as
\al{
\gamma_m=\frac{1}{16\pi^2}\left( 2\lambda + 6y_t^2 -\frac{9}{2} g_2^2 -\frac{3}{2} g_1^2 \right)\,.
}
This value at the electroweak scale is $\gamma_m\approx 0.027$, so that it is negligible in comparison with the canonical scaling $2$ in the RG equation for the scalar mass \eqref{perturbative mass RG equation}.
We here evaluate the value of $\bar m_0$ when a bare theory of the SM is given at the Planck scale $\Lambda=M_\text{pl}\simeq 10^{19}$\,GeV using Eq.\,\eqref{running scalar mass} with $\gamma_m=0$ .
The observed Higgs boson mass $m_H=125$\,GeV and the electroweak scale $v_h=246$\,GeV tell us that the dimensionless renormalized mass at the electroweak scale is given as $\bar m^2(\mu=v_h)=m_H^2/v_h^2\simeq 0.1$\,.
From this one finds $\bar m^2(\mu=M_\text{pl})=m_0^2/M^2_\text{pl}\simeq 10^{-36}$ at the Planck scale.
Hence, the scalar mass in the bare action is quite smaller than the Planck scale at which the bare action is defined.
This is the remaining gauge hierarchy problem.

This gauge hierarchy problem originates from the fact that the scalar mass term is a relevant parameter with the large critical exponent $\theta_m\approx 2$.
Therefore, one of simple possible solutions to this problem is to have a large anomalous dimension $\gamma_m$ such that $\theta_m=2-\gamma_m \lesssim 0$; see e.g.~\cite{Wetterich:1981ir,Bornholdt:1992up,Terao:2007jm,Wetterich:2011aa}.
To realize such a situation, we need strong dynamics.
However, new particles strongly coupled to the Higgs field around the TeV scale are highly constrained by the collider experiments such as the LHC.

An other possibility is classical scale invariance~\cite{Wetterich:1983bi,Bardeen:1995kv}.
Classical scale symmetry forbids the renormalized dimensionful mass $m^2$ at the UV scale $\Lambda$.
Hence it enforces, for the bare theory,
\al{
\widehat m_0^2=0\,.
}
In this case, the renormalized scalar mass keeps vanishing under varying the RG scale as can be seen in Eq.\,\eqref{running scalar mass}.
This fact means that the breaking of scale symmetry due to the scale anomaly is not soft, but is hard.
On the other hand, how does classical scale symmetry act on $\delta m_0^2$ which is also a dimensionful parameter?
Scale symmetry is broken by the quadratic divergence coming from regularizations such as the Pauli-Villars regularization and the momentum regularization.
Such an explicit breaking should be modified in accordance with the Ward-Takahashi identity for scale symmetry, i.e. the Callan-Symanzik equation~\cite{Symanzik:1970rt,Callan:1970yg}.
Hence, one can interpret $\delta m_0^2$ as an additive counter term in order to remove the spurious breaking of scale symmetry. 
Needless to say, the dimensional regularization respects to scale symmetry and then yields $\delta m_0^2=0$.
Then the massless scalar theory is renormalizable~\cite{Lowenstein:1975rf}.

In the viewpoint from the Wilsonian renormalization group, the massless theory is realized at the phase boundary (critical line).
The quite small Higgs mass $\widehat m_0/M_\text{pl}\simeq 10^{-36}$ means that the bare Higgs is located very near the phase boundary.
In this sense, the gauge hierarchy problem can be paraphrased as {\it the criticality problem}: {\it Why is the Higgs so close to critical?}
Classical scale symmetry forces the Higgs to be put exactly on the phase boundary: {\it Classical scale symmetry makes the Higgs critical.}

The theory with classical scale invariance has no corresponding scale to the electroweak one, so that one needs to a mechanism in order to generate a scale, i.e. scalagenesis is required. 
A simple mechanism for scalegenesis is the Coleman-Weinberg mechanism.
This is, however, not compatible with the observed masses of SM particles in order to generate the electroweak scale.
Therefore, we need an extension of the SM for the use of the Coleman-Weinberg mechanism.
 We discuss details in Section~\ref{Section: Scalegenesis}.

Within low energy theories, however, we cannot explain why the SM or its extended models becomes scale invariant at a certain UV scale or the Planck scale.
This is nothing but a strong assumption.
It is interesting whether or not the scale invariant effective theory at the Planck scale is naturally defined from UV theories including quantum gravity.
We will discuss this issue within the asymptotic safety scenario for quantum gravity in Section~\ref{Sec: Asymptotically safe quantum gravity and gauge hierarchy problem}.
In particular, we will see that a large anomalous dimension could be induced by graviton fluctuations above the Planck scale and thus it could provide natural solutions to the gauge hierarchy problem.
In particular, classical scale invariance at the Planck scale emerges as a boundary condition from asymptotically safe quantum gravity.

\section{Scalegenesis}
\label{Section: Scalegenesis}
If we impose classical scale symmetry on the SM, there is no corresponding scale to the electroweak one.
Therefore, we need a mechanism to generate a scale. 
There are two ways for scalegenesis: the one is the Coleman-Weinberg mechanism~\cite{Coleman:1973jx} in the perturbation theory; and the other relies on the strong dynamics like quantum chromodynamics.
The next section is devoted to discussing the basis mechanism of scalegenesis.
We might think that the most minimal way to generate the electroweak scale is the Coleman-Weinberg mechanism within the SM.
Unfortunately, this does not work since the observed SM particle masses do not satisfy the condition for scalegenesis.
We see this fact in Section~\ref{Sec: Coleman-Weinberg mechanism in the SM}.
Hence, we need an extension of the SM.
In Section~\ref{Sec: A model for scalegenesis}, as an example of scalegenesis, we introduce an extended model of the SM where a strongly interacting scalar-gauge theory in a hidden section is given.

\subsection{Condition for scalegenesis}
We review the Coleman-Weinberg mechanism bases on the original paper~\cite{Coleman:1973jx} to understand the crucial point for the generation of a scale.
We consider a U(1) scalar-gauge theory whose action is given by
\al{
S=\int \df^4 x\left[ \frac{1}{2}(D_\mu \phi)^2 -\frac{\lambda}{4!}\phi^4 -\frac{1}{4} F_{\mu\nu}F^{\mu\nu}\right]\,,
}
where $\phi$ is a complex scalar field, $F_{\mu\nu}$ is the field strength of $A_\mu$, $D_\mu =\p_\mu -ie A_\mu$ is the covariant derivative.
The effective potential for the scalar field at the one-loop level is computed as
\al{
V_{\rm eff}\fn{\phi}= \frac{\lambda}{4!}\phi^4 +\left( \frac{5\lambda^2}{18}+ 3e^4\right)\frac{\phi^4}{64\pi^2}\left( \ln \frac{\phi^2}{v_\phi^2} - \frac{25}{6}\right) :=\frac{\lambda_{\text{eff}}(\phi)}{4!}\phi^4\,,
}
where $v_\phi=\langle \phi\rangle$.
The vacuum condition is obtained by the first derivative of the effective potential with respect to $\phi$, namely
\al{
&\phi\frac{\df V_{\rm eff}\fn{\phi} }{\df \phi}\bigg|_{\phi=v_\phi}=0 \qquad
 \Longleftrightarrow \qquad
\left[4\lambda_{\text{eff}}(\phi) + \beta_\lambda \right] \big|_{\phi=v_\phi}=0\,,
\label{Eq: vacuum condition}
}
where $\beta_\lambda = \phi \frac{\df  \lambda_{\rm eff}\fn{\phi}}{\df \phi}$ in the beta function of the quartic coupling.
From this we find, for a small $\lambda$,
\al{
\left( \lambda - \frac{33e^4}{8\pi^2} \right) v_\phi=0\,.
}
We see from this that the requirement of a non-vanishing vacuum $v_\phi\neq 0$ gives a relation between coupling constants: $\lambda=33e^4/8\pi^2$.
Thus, one of two dimensionless couplings ($\lambda$ and $e$) changes to the dimensionful quantity, i.e. $v_\phi$.
Coleman and Weinberg called this mechanism the dimensional transmutation.
Recently, the author and his collaborator have suggested calling it scalegenesis~\cite{Kubo:2015cna}.

The crucial point for the generation of a scale in the Coleman-Weinberg mechanism is that the effective quartic coupling constant becomes negative for small field values.
In this simple model, the gauge coupling plays a central role for this: In terms of the renormalization group equation, one has the loop effect of the gauge field on the beta function of the scalar quartic coupling such that $\beta_\lambda \supset +e^4$.
That is, the effect of the gauge coupling makes the RG flow of the quartic coupling negative towards the IR  (small field value) region.
An other possible model of the Coleman-Weinberg mechanism is the system with the three scalar fields, e.g. the Higgs field plus two additional singlet-scalar fields~\cite{Haruna:2019zeu}.
The coupling constants $\lambda_{ij}$ between two different scalar fields such as the Higgs portal coupling contribute to the beta functions of quartic couplings $\lambda_i$ so that $\beta_{\lambda_i}\supset +\lambda_{ij}^2$ where indices denote species of scalar fields.

We here note that in the perturbative scalegenesis, the scalar mass parameter, which corresponds to the curvature at the origin of the effective potential, vanishes.
This reflects the fact that the scale anomaly does not induce the soft breaking of scale symmetry as mentioned in the previous section.
On the other hand, in scalegenesis due to non-perturbative dynamics, scale symmetry is spontaneously broken and the scalar field obtains a dynamical (constituent) mass $M^2$.
In particular, we observe a divergence of the quartic coupling at a certain scale which is an origin of a scale.
We see this fact in Section~\ref{Sec: A model for scalegenesis}.

\subsection{Effective couplings}
An interesting question is how we can experimentally distinguish scalegensis by the Coleman-Weinberg mechanism from the SM type due to a negative scalar mass parameter.
Indeed, between them, there are large differences of the Higgs effective couplings which are defined as
\al{
&\lambda^{(2)}= \frac{1}{2v_h^2}\frac{\df^2V}{\df h^2}\bigg|_{h=v_h}\,,&
&\lambda^{(3)}= \frac{1}{6v_h}\frac{\df^3V}{\df h^3}\bigg|_{h=v_h}\,,&
&\lambda^{(4)}= \frac{1}{6}\frac{\df^4V}{\df h^4}\bigg|_{h=v_h}\,.&
\label{Eq: Higgs effective couplings}
}
Let us see explicitly these couplings for both the SM and the Coleman-Weinberg cases.
In the SM case in which the Higgs potential with $v_h=\sqrt{m^2/\lambda}$ is given by
\al{
V_\text{SM}(h)=-\frac{m^2}{2} h^2 +\frac{\lambda}{4}h^4\,,
}
the Higgs effective couplings \eqref{Eq: Higgs effective couplings} become $\lambda_\text{SM}^{(2)}=\lambda_\text{SM}^{(3)}=\lambda_\text{SM}^{(4)}=\lambda$ which leads to
\al{
&\frac{\lambda_\text{SM}^{(3)}}{\lambda_\text{SM}^{(2)}}= 1\,,& 
&\frac{\lambda_\text{SM}^{(4)}}{\lambda_\text{SM}^{(2)}} =1\,.
\label{Eq: SM effective couplings}
}
We next suppose that an extension of the SM realizes electroweak scalegenesis by the Coleman-Weinberg mechanism in the Higgs sector and gives the following effective potential,
\al{
V_\text{CW}(h)= \frac{\lambda_\text{eff}(h)}{4}h^4\,.
}
For this potential, we obtain the vacuum from Eq.\,\eqref{Eq: vacuum condition} and find 
\al{
\lambda_\text{CW}^{(2)}&= \frac{1}{2}\left( \lambda_\text{eff}' +\frac{\lambda_\text{eff}''}{4}\right)\,, \\[2ex]
\lambda_\text{CW}^{(3)}&= \frac{5}{3}\lambda_\text{CW}^{(2)} +\frac{\lambda_\text{eff}''}{6} + \frac{\lambda_\text{eff}'''}{24} \,,
\label{EC 3}
\\[2ex]
\lambda_\text{CW}^{(4)}&= \frac{11}{3}\lambda_\text{CW}^{(2)} + \lambda_\text{eff}'' +\frac{5\lambda_\text{eff}'''}{12} + \frac{\lambda_\text{eff}''''}{24} \,,
\label{EC 4}
}
where a prime denotes the derivative $h\frac{\df}{\df h}$.
Then, neglecting the higher order terms in Eqs.\,\eqref{EC 3} and \eqref{EC 4}, we find the ratios~\cite{Dermisek:2013pta}
\al{
&\frac{\lambda_\text{CW}^{(3)}}{\lambda_\text{CW}^{(2)}}=\frac{5}{3}\,,&
&\frac{\lambda_\text{CW}^{(4)}}{\lambda_\text{CW}^{(2)}}=\frac{11}{3}\,.
\label{Eq: CW effective couplings}
}
We see that the Higgs effective couplings \eqref{Eq: CW effective couplings} in the Coleman-Weinberg potential drastically deviate from the prediction from the SM \eqref{Eq: SM effective couplings}.
The collider experiments such as the international linear collider in future will measure the Higgs effective coupling precisely and clarify the origin of electroweak scalegenesis.

\subsection{Coleman-Weinberg mechanism in the SM}
\label{Sec: Coleman-Weinberg mechanism in the SM}
We briefly investigate at the Coleman-Weinberg potential in the SM and see that the electroweak scalegenesis actually does not take place.
We have the one-loop effective potential in the SM,
\al{
V_{\rm eff}\fn{h} =  \frac{\lambda_H}{4}h^4+
\sum_\alpha \frac{N_\alpha M_\alpha^4(h)}{64\pi^2}\left( \ln \frac{M_\alpha^2(h)}{v_h^2}  -C_\alpha \right)
=: \frac{\lambda_{H,\,\text{eff}}(h)}{4}h^4\,,
\label{Eq: one-loop effective potential in the SM}
}
where $\alpha=(W,\,Z,\,t,h)$; $N_W=6$, $N_Z=3$, $N_t=-12$ and $N_h=1$; and $C_{h,t}=3/2$ and $C_{W,Z}=5/6$; and the masses at the tree level are given by
\al{
&M_W^2= \frac{g_2^2}{4}h^2\,,&
&M_Z^2=\frac{g_2^2+g_1^2}{4}h^2\,,&
&M_t^2= \frac{y_{t}^2 }{2}h^2\,,
&M_h^2= 3\lambda_H h^2\,.
}
Here, $\lambda_{H,\,\text{eff}}(h)$ is the effective quartic coupling as a functions of $h$.
We have neglected contributions from other SM particles in the effective potential \eqref{Eq: one-loop effective potential in the SM}.
The vacuum condition \eqref{Eq: vacuum condition} in the present case yields 
\al{
v_h^2=\frac{-36 M_t^4+6 M_W^4+3M_Z^4+3M_h^4}{16 \pi^2 M_h^2}\,.
\label{Eq: vacuum condition in SM}
}
For the observed masses, $M_W=80$\,GeV, $M_Z=91$\,GeV\,, $M_t=171$\,GeV and $M_h=125$\,GeV, the relation \eqref{Eq: vacuum condition in SM} yields a negative value of $v_h^2$.
Therefore, even a finite value of the vacuum cannot be generated.
This is because the Higgs quartic coupling does not become negative for small field values.
If the top-quark mass was smaller or the Higgs and gauge boson masses were larger, a finite scale would be generated by the Coleman-Weinberg mechanism.

\subsection{Scale from new physics}
\label{Sec: Scale from new physics}
As we have seen in Section~\ref{Sec: Coleman-Weinberg mechanism in the SM}, the Coleman-Weinberg mechanism in the SM does not occur
Therefore, we need an extension of the Higgs sector.
The simplest extension may be an introduction of a scalar field $S$ coupled to the Higgs field through the Higgs-portal coupling.
We give its Lagrangian as
\al{
\mathcal L= \mathcal L_\text{SM}|_{m_H\to0} - \lambda_{HS}(H^\dagger H) (S^\dagger S) + \mathcal L_{S}|_{m_S\to 0} \,,
}
where $\mathcal L_{S}|_{m_S\to 0}$ is the Lagrangian for the scalar field $S$ and we call it a hidden sector.
We do not here specify the explicit form of $\mathcal L_{S}|_{m_S\to 0}$.
Assuming that the Coleman-Weinberg mechanism or the strong dynamics in the hidden sector works and then we obtain a non-trivial vacuum $\langle S\rangle \neq 0$ or $\langle S^\dagger S\rangle \neq0$ which is proportional to a scale $\Lambda_H$ (smaller than the Planck scale), the Higgs-portal coupling plays a role of the Higgs mass term
\al{
m_H^2 \simeq \lambda_{HS} \Lambda_H^2\,.
\label{Eq: Higgs mass generation in general}
}
In order for the Higgs field to obtain a non-trivial vacuum, the Higgs-portal coupling has to be negative.
Moreover, $\Lambda_H$ has to be of order of about TeV scale if the Higgs-portal coupling is $\mathcal O(10^{-3})$--$\mathcal O(10^{0})$.

It seems, however, that the Higgs mass parameter in Eq.\,\eqref{Eq: Higgs mass generation in general} receives a contribution from the quadratic divergence.
Is it then irrelevant for the Higgs mass term by recalling the discussion in Section~\ref{Sec: Gauge hierarchy problem in the SM and classical scale invariance}?
Its answer is No.
Indeed, the quadratic divergence in Eq.\,\eqref{Eq: Higgs mass generation in general} is physical.
In order to clarify this, we discuss the Wilsonian RG for a simple system with two scalar fields,
\al{
\Gamma_k[\phi] \simeq \int \df^4 x \left[ \frac{1}{2}(\p_\mu \phi_i)^2 + \frac{1}{2}m_{i}^2  \phi_i^2  + \frac{1}{4!}\lambda_{i} \phi_i^4 +\frac{1}{8}\lambda_{ij} \phi_i^2 \phi_j^2  \right]\,,
}
where $i=1,\,2$, $i\neq j$, $\lambda_{12}=\lambda_{21}$ and we assume that $m_1<m_2$.
Using the flow equation \eqref{Wetterich equation}, we obtain the RG equations,
\al{
\p_t \bar m_{i}^2 &= -2{\bar m}_{i}^2 - \frac{\bar \lambda_{i}}{32\pi^2} \ell_1^4({\bar m}_i^2)- \frac{\bar \lambda_{ij}}{32\pi^2}\ell_1^4({\bar m}_j^2)\nn[1ex]
&\simeq -2{\bar m}_{i}^2 - \frac{\bar \lambda_{i}+\bar \lambda_{ij}}{32\pi^2}  \ell_1^4(0) + \frac{\bar \lambda_{i}}{16\pi^2} \ell_2^4(0){\bar m}_i^2 + \frac{\bar \lambda_{ij}}{16\pi^2} \ell_2^4(0){\bar m}_j^2 \,,
 \label{Eq: two scalar system beta functions of masses}
 \\[2ex]
\p_t \bar\lambda_{i}&=  -\frac{3\bar \lambda_{i}^2}{16\pi^2}\ell_2^4({\bar m}_i^2)
- \frac{3\bar \lambda_{ij}^2}{16\pi^2}\ell_2^4({\bar m}_j^2)\\[2ex]
\p_t \bar\lambda_{ij}&=  - \frac{\bar \lambda_{ij}^2}{8\pi^2}\left[\ell_0^4({\bar m}_i^2) \ell_1^4({\bar m}_j^2)+\ell_0^4({\bar m}_j^2) \ell_1^4({\bar m}_i^2) \right]
+\frac{\bar\lambda_{ij}}{16\pi^2}\left[ \bar\lambda_i\ell_2^4({\bar m}_i^2)+\bar\lambda_j\ell_2^4({\bar m}_j^2)  \right]
\,,
}
In the beta functions of the scalar masses \eqref{Eq: two scalar system beta functions of masses}, we expand the threshold functions into polynomials of $\bar m_i^2$ and take into account of order of $\bar m_i^2$ by assuming $\bar m_i^2 \ll 1$. 
Here we analyze the RG flows around the Gaussian fixed point.
To see a weak mixing effect between the two scalar fields, we take into account a linear order of $\bar\lambda_{ij}$ in Eq.\,\eqref{Eq: two scalar system beta functions of masses}.
The linearized RG equations around the Gaussian fixed point $(\bar m^2_{i*}, \bar\lambda_{i*})=(0,0)$ and $\bar\lambda_{ij}\ll 1$ are given by
\al{
\p_t \pmat{
\bar m_1^2\\[1ex]
\bar m_2^2\\[1ex]
\bar \lambda_1\\[1ex]
\bar \lambda_2\\[1ex]
\bar\lambda_{12}
}
 \simeq  
\pmat{
-2 &&  \frac{\bar\lambda_{12}}{16\pi^2} && -\frac{a}{32\pi^2} && 0 && -\frac{a}{32\pi^2} \\[2ex]
\frac{\bar\lambda_{12}}{16\pi^2} && -2 && 0 &&  -\frac{a}{32\pi^2} &&  -\frac{a}{32\pi^2} \\[2ex]
0 && 0 && 0 && 0 && 0 \\[2ex]
0 && 0 && 0 && 0 && 0 \\[2ex]
0 && 0 && 0 && 0 && 0 
}
\pmat{
\bar m_1^2\\[1ex]
\bar m_2^2\\[1ex]
\bar \lambda_1\\[1ex]
\bar \lambda_2\\[1ex]
\bar\lambda_{12}
}\,,
\label{linearized equation in two scalar theory}
}
where we have used the Litim-type cutoff \eqref{Eq: Litim cutoff} to have $\ell_1^4(0)=a$ and $\ell_2^4(0)=1$.
The stability matrix can be diagonalized by the rotation matrix,
\al{
V\simeq \pmat{
1 && -1 &&-\frac{a}{64\pi^2}   && \frac{a}{64\pi^2}\frac{\bar\lambda_{12}}{32\pi^2} &&  -\frac{a}{64\pi^2}\left(1 -\frac{\lambda_{12}}{32\pi^2}\right)\\[1ex]
1 && 1 &&  \frac{a}{64\pi^2}\frac{\bar\lambda_{12}}{32\pi^2} && -\frac{a}{64\pi^2}  && -\frac{a}{64\pi^2}\left(1 -\frac{\lambda_{12}}{32\pi^2}\right) \\[1ex]
0 && 0  && 1 && 0 && 0  \\[1ex]
0 && 0 &&  0 &&  1 && 0 \\[1ex]
0 && 0 &&  0 &&  0 && 1
}\,,
\label{rotation matrix in an example 2}
} 
and has the eigenvalues, i.e. the critical exponents,
\al{
(\theta_{m_1},\theta_{m_2},\theta_{\lambda_1},\theta_{\lambda_2},\theta_{\lambda_{12}})\simeq \left(-2+\frac{\bar\lambda_{12}}{16\pi^2},\,-2-\frac{\bar\lambda_{12}}{16\pi^2},\, 0,\,0,\,0 \right)\,.
}
Then the RG equations for the scalar masses reads
\al{
&\p_t \widetilde m_1^2 = -2\widetilde m_1^2 + \frac{\bar\lambda_{12}}{16\pi^2}\widetilde m_2^2 \,, 
\label{Eq: RG eq for m1}
\\[2ex]
&\p_t \widetilde m_2^2 = -2\widetilde m_2^2 \,,
\label{Eq: RG eq for m2}
}
where we have defined
\al{
&\widetilde m_1^2= \bar m_1^2  + \frac{a\bar\lambda_1}{64\pi^2}
+ \frac{a \bar\lambda_{12}}{64\pi^2}\left(1 + \frac{\bar\lambda_{12}}{32\pi^2}+\frac{\bar\lambda_{2}}{32\pi^2}\right) 
 \,,\\[2ex]
&\widetilde m_2^2=  \bar m_2^2 + \frac{a\bar\lambda_2}{64\pi^2}
+ \frac{a \bar\lambda_{12}}{64\pi^2}\left(1 + \frac{\bar\lambda_{12}}{32\pi^2}+\frac{\bar\lambda_{1}}{32\pi^2}\right) \,. 
}
In Eqs.\,\eqref{Eq: RG eq for m1} and \eqref{Eq: RG eq for m2}, the quadratic divergent terms depending on the cutoff scheme are subtracted.
Let us now solve the RG equations.
The solution to Eq.\,\eqref{Eq: RG eq for m2} is easily found to be
\al{
\widetilde m_2^2(k) 
 =\widetilde  m_{2,0}^2  \left( \frac{k}{\Lambda_2}\right)^{-2}\,,
}
where $\Lambda_2$ is a scale at which the mass of $\phi_2$ is given, e.g. the expectation value of $\phi_2$, and $\widetilde  m_{2,0}$ is a boundary value of $\widetilde m_2$ at $k=\Lambda_2$.
The avoidance of the scale hierarchy in the $\phi_2$ sector requires $\widetilde  m_{2,0}\simeq 1$.
Substituting this solution into Eq.\,\eqref{Eq: RG eq for m1} and ignoring the running of $\bar\lambda_{12}$, we find 
\al{
\widetilde m_1^2(k) =\widetilde m_{1,0}^2 \left( \frac{k}{\Lambda_2} \right)^{-2} +\frac{\bar\lambda_{12}}{16\pi^2} \widetilde  m_{2,0}^2 \left( \frac{k}{\Lambda_2} \right)^{-2} \ln \fn{\frac{k}{\Lambda_2}}\,.
}
We see from this equation that even if $\widetilde m_{1,0}^2=0$ the scalar field $\phi_1$ obtains a finite dimensionful mass
\al{
\widehat m_1^2(\Lambda_1) = \frac{\bar\lambda_{12}}{16\pi^2} \widehat  m_{2,0}^2(\Lambda_2)\ln \fn{\frac{\Lambda_1}{\Lambda_2}} \simeq \bar\lambda_{12} \Lambda_2^2 \,,
\label{Eq: m1 solution}
}
where $\Lambda_1<\Lambda_2$, we have defined $\widehat m_1^2(\Lambda_1) =\widetilde m_1^2(\Lambda_1) \Lambda_1^2$ and $\widehat m_2^2(\Lambda_2) =\widetilde m_2^2(\Lambda_2) \Lambda_2^2\simeq \Lambda_2^2$ and have assumed that $\ln \fn{{\Lambda_1}/{\Lambda_2}}$ is of order one.
This is just Eq.\,\eqref{Eq: Higgs mass generation in general}.
We conclude that the quadratic divergence in Eq.\,\eqref{Eq: Higgs mass generation in general} or Eq.\,\eqref{Eq: m1 solution} is not be subtracted by rotating the theory space and then is a physical object independent from the cutoff scheme.

\subsection{A model for scalegenesis}
\label{Sec: A model for scalegenesis} 
We show one of examples for scalegenesis owing to the strong dynamics of the gauge interaction~\cite{Kubo:2015cna,Kubo:2014ova}.
We introduce a scale invariant hidden sector in which a scalar field is coupled to the SU$(N_c)$ gauge field.
The scalar field is also coupled to the Higgs field via the Higgs-portal coupling. 
The total Lagrangian reads
\al{
\mathcal L= \mathcal L_\text{SM}|_{m_H=0} + \mathcal L_\text{hidden}\,.
\label{Eq: Model lagrangian}
}
Here the Lagrangian for the hidden sector is given by
\al{
 \mathcal L_\text{hidden}&=-\frac{1}{2}~\mbox{tr} \,F^2
+([D^\mu S_i]^\dag D_\mu S_i)
-\hat{\lambda}_{S}(S_i^\dag S_i) (S_j^\dag S_j)
-\hat{\lambda}'_{S}
(S_i^\dag S_j)(S_j^\dag S_i)
+\hat{\lambda}_{HS}(S_i^\dag S_i)H^\dag H\,,
\label{Eq: hidden Lagrangian}
} 
where $F=F^a \tau^a$ is the field strength of SU$(N_c)$ gauge field $A_\mu^a$; $\tau^a$ is the generator of SU$(N_c)$ gauge transformation; $S_i$ and $H$ are the new scalar field and the Higgs doublet field, respectively; indices on the scalar field $S$ stand for the flavor indices; and $D_\mu=\p_\mu - ig A_\mu^a t^a$ is the covariant derivative.
This hidden Lagrangian \eqref{Eq: hidden Lagrangian} is invariant under scaler symmetry, the SU$(N_c)$ gauge symmetry and the U$(N_f)$ flavor symmetry.
Note that $\hat{\lambda}_{HS}$ takes a positive value and the Higgs quartic interaction is included in $\mathcal L_\text{SM}|_{m_H=0}$.

We assume that the SU$(N_c)$ gauge symmetry is not broken by the dynamics in the hidden sector, but scale symmetry is spontaneosely broken.
Due to the strong dynamics of the gauge field in the low energy region the SU$(N_c)$ invariant scalar bilinear condensate takes place such that
\al{
\langle S^\dag_i S_j\rangle =
\left\langle \sum_{a=1}^{N_c} S^{a\dag}_i S^a_j \right\rangle\propto \delta_{ij}\,.
\label{Eq: bilinear condensate}
}
Thus, the Higgs portal coupling takes a form of the (negative) Higgs mass parameter: 
\al{
m_H^2 = -\hat\lambda_{HS} \langle S^\dag_i S_i \rangle \,.
} 
As a consequence, the Higgs field has a non-trivial vacuum $v_h=\sqrt{m_H^2 /\lambda_H}$.

We would like to see that the scalegenesis introduced above actually takes place.
It is, however, quite difficult to analyze the vacuum structure in the original Lagrangian \eqref{Eq: Model lagrangian}.
Instead, we here attempt to formulate an effective model which describes the dynamical scale symmetry breaking, {\it {\` a} la}, the Nambu--Jona-Lasinio model as a chiral effective model of quantum chromodynamics.
However, scale symmetry is broken by not only the bilinear condensation \eqref{Eq: bilinear condensate}, but also the scale anomaly.
Nevertheless, the latter breaking effect is hard, namely the breaking of scale symmetry is due to higher dimensional operators as discussed in the previous section.
Therefore, we could ignore anomalous breaking effects in low energy regions.
Under this assumption, we employ the following effective Lagrangian to describe the spontaneous scale symmetry breaking~\cite{Kubo:2015cna}:\footnote{
The confinement effect in terms of the Polyakov loop is discussed in Ref.\,\cite{Kubo:2018vdw}.
}
\al{
  {\cal L}_{\rm eff} &=
 ([\partial^\mu S_i]^\dag \partial_\mu S_i)-
\lambda_{S}(S_i^\dag S_i) (S_j^\dag S_j)
-\lambda'_{S}
(S_i^\dag S_j)(S_j^\dag S_i)
 +\lambda_{HS}(S_i^\dag S_i)H^\dag H
-\lambda_H ( H^\dag H)^2\,,
\label{Eq: effective Lagrangian}
}
where we include the Higgs quartic coupling.\
It is supposed in the effective Lagrangian \eqref{Eq: effective Lagrangian} that quantum fluctuations of the gauge field $A_\mu$ were integrated out and then the scalar fields and the coupling constants are defined as effective ones.

Using the auxiliary field method in the path integral formalism or the normal ordering method in the operator formalism, we can obtain the mean-field approximated effective Lagrangian so that
\al{
{\cal L}_\text{MFA}&=
 ([\partial^\mu S_i]^\dag \partial_\mu S_i) 
 -M^2(S_i^\dag S_i)
 -\lambda_H ( H^\dag H)^2  
 + N_f (N_f\lambda_S+\lambda'_S) f^2 
+\frac{\lambda_S'}{2}(\phi^a)^2
-2\lambda_S' \phi^a(S_i^\dagger t^a_{ij} S_j),
\label{Eq: mean-field approximated Lagrangian}
}  
where $f=(S_i^\dagger S_i)/N_f$ and $\phi^a=2(S_i^\dagger t^a_{ij} S_j)$ are auxiliary fields with $t^a$ the generator of the flavor SU$(N_f)$ transformation, and the ``constituent" scalar mass is given by
\al{
M^2 = 2(N_f\lambda_S+\lambda'_S)f -\lambda_{HS} H^\dagger H.
\label{cons scala mass}
}

Assuming that the bilinear condensate \eqref{Eq: bilinear condensate} is invariant under the U$(N_f)$ flavor transformation, we can choose a vacuum state $\langle f \rangle \neq 0$ and $\langle \phi^a \rangle =0$.
Then, we can set to $\phi^a=0$ in Eq.\,\eqref{Eq: mean-field approximated Lagrangian} in order to derive the effective potential.
The scalar field $S_i$ takes the bilinear form in the Lagrangian \eqref{Eq: mean-field approximated Lagrangian}, so that we can integrate out it and obtain the effective potential,
\al{
V_\text{MFA}\fn{\bar S,f,H}
&= M^2 (\bar S_i^\dagger \bar S_i) +\lambda_H (H^\dagger H)^2 -N_f (N_f \lambda_S +\lambda'_S)f^2 +\frac{N_cN_f}{32\pi^2}M^4 \ln\frac{M^2}{\Lambda_H^2}\,,
\label{effective potential MFAa}
}
where $\bar S_i$ is the background field of the scalar field $S_i$, we have employed the dimensional regularization and the $\overline{\text{MS}}$ scheme to subtract a UV divergence; and $\Lambda_H$ is a renormalization point at which the quantum effect vanishes for $M=\Lambda_H$.

To find the vacuum of the system, we evaluate the gap equations, i.e. the stationary conditions, 
\al{
0=\frac{\del}{\del \bar S^a_i}V_{\rm MFA}
= \frac{\del}{\del f}V_{\rm MFA}
=\frac{\del}{\del H_l}V_{\rm MFA}\qquad(l=1,2)\,,
\label{station}
}
from which we find a minimum\footnote{
In general, one can consider three possibilities as a vacuum solution: (i)~$\langle \bar S^a_i\rangle\neq 0$ and $\langle f \rangle=0$; (ii)~$\langle \bar S^a_i\rangle= 0$ and $\langle f \rangle=0$; (iii)~$\langle \bar S^a_i\rangle= 0$ and $\langle f \rangle \neq 0$.
The last condition (iii) yields absolute minimum of the effective potential.
}
\al{
&\langle \bar S_i \rangle =0\,,&
&\langle f\rangle =f_0=\frac{2 \lambda_H}{G} 
\Lambda_H^2\exp\left(  \frac{32\pi^2 \lambda_H}{N_c G}-\frac{1}{2}\right)\,,
 \label{vev1}&
&|\langle H\rangle |^2
 =\frac{v_h^2}{2}=
 \frac{N_f\lambda_{HS}}{2\lambda_H}\langle f\rangle \,,
}
where $G=4N_f \lambda_H \lambda_S-N_f \lambda_{HS}^2+4 \lambda_H\lambda'_S$.
At this vacuum the constituent scalar mass and the Higgs mass are given by
\al{
 &\langle M^2\rangle = M_0^2=\frac{G}{2\lambda_H} \langle f\rangle \,,
& 
&  M_{h}^2= |\langle H\rangle |^2\left(
\frac{16\lambda_H^2
(N_f\lambda_S+\lambda'_S)}{G}
+\frac{N_c N_f
\lambda_{HS}^2}{8\pi^2}\right) \simeq 2N_f\lambda_{HS} \langle f\rangle \,,
\label{vev2}
}
where we assumed a small $\lambda_{HS}$ in the Higgs mass.
We see that the dimensionful quantity $\Lambda_H$ is generated at the quantum level and becomes an origin of the electroweak scale and the Higgs mass.

Let us here the relation between the generation of the scale $\Lambda_H$ and the behavior of the quartic coupling $\lambda_S$ in the hidden sector. 
For simplicity, we consider the $N_f=1$ case and set $\lambda_{HS}=0$.
We have the effective potential for $\chi:=2\lambda_S f$,
\al{
V_{\rm MFA}\fn{\chi; \lambda_S} = -\frac{1}{4\lambda_S} \chi^2 +\frac{N_c}{32\pi^2}\chi^2 \ln \frac{\chi}{\Lambda_H^2}\,.
\label{Eq: MFA potential for chi}
}
From 
\al{
\frac{\p ^2 V_{\rm MFA}}{\p \chi^2}\fn{e^t \chi; \bar\lambda_S}
=\frac{\p ^2 V_{\rm MFA}}{\p \chi^2}\fn{\chi;\lambda_S\fn{t}}\,,
}
the RG flow of the quartic coupling $\lambda_S$ reads
\al{
\lambda_S\fn{t}=  \frac{\bar\lambda_S}{1-\displaystyle\frac{N_c \bar\lambda_S}{8\pi^2}t}\,,
}
where $t$ is a dimensionless scaling parameter and we defined $\lambda_S(t=0)=\bar\lambda_S>0$.
Obviously, there is a Landau pole at $t=t_c=8\pi^2/N_c \bar\lambda_S$.
Since the present analysis does not rely on the perturbative expansion of $\lambda_S$, the system is still defined for $t>t_c$ for which the effective quartic coupling becomes negative.
The minimum of the effective potential \eqref{Eq: MFA potential for chi} is located at
\al{
\chi_{\rm m}:=  \Lambda_H^2\exp\left(\frac{8\pi^2}{\Nc \bar\lambda_S} -\frac{1}{2} \right)\,.
\label{Eq: minimum of chi}
}
The corresponding dimensionless scaling parameter to this minimum is
\al{
t=t_{\rm m}:=\ln\fn{\frac{\chi_{\rm m}}{\mu^2}} = \left( \frac{8\pi^2}{N_c \bar \lambda_S} -\frac{1}{2}  \right) + \ln\frac{\Lambda_H^2}{\mu^2}
= t_c -\frac{1}{2} + \ln\frac{\Lambda_H^2}{\mu^2}\,,
}
where $\mu$ is a dimensionful scale at which $\bar\lambda_S$ is given.
In particular, for the choice\footnote{
This choice of $\mu$ is equivalent to the redefinition of $\lambda_S(t=0)=\bar\lambda_S$ with the choice $\mu=\Lambda_H$ such that $\frac{8\pi^2}{\Nc \bar\lambda_S} -\frac{1}{2} \to \frac{8\pi^2}{\Nc \bar\lambda_S}$ in Eq.\,\eqref{Eq: minimum of chi}.
This is nothing but the dimensional transmutation or scalegensis: the degree of freedom of the dimensionless coupling $\bar\lambda_S$ changes to that of  the dimensionful parameter.
} $\mu=\Lambda_H e^{-1/4}$, we have $t_\text{m}=t_c$.
We see that the non-trivial vacuum \eqref{Eq: minimum of chi} is related to the Landau pole of the effective quartic coupling $\lambda_S(t)$.

We note here that for $\chi\to \infty$ the quadratic coupling goes to zero from the negative side of $\lambda_S(t)$, i.e. it is asymptotically free. 
In such a case, however, the potential of the scalar field $S$ is unbounded, so that $S$ is unstable.
This situation could be improved by taking account the dynamics of the gauge field.
We will discuss this issue elsewhere.

We finally comment on several phenomenological implications from this model.
The composite scalars $\phi^a$ are stable due to the flavor symmetry, so that they can be dark matter candidates.
The prediction for the spin-independent elastic cross section of $\phi^a$ off the nucleon could be tested by the direct detection experiments~\cite{Kubo:2015cna,Kubo:2017wbv}.
In this model, the scale and electroweak phase transitions take place at finite temperature.
In particular, the scale phase transition becomes the strong first-order, i.e. $\langle f\rangle^{1/2}/T_c \gtrsim 1$ where $T_c$ is the critical temperature~\cite{Kubo:2015joa}.
The strong first-order scale phase transition produces gravitational waves which could be observed by the future space gravitational wave
antennas~\cite{Kubo:2016kpb}.

\section{Asymptotically safe quantum gravity and the gauge hierarchy problem}
\label{Sec: Asymptotically safe quantum gravity and gauge hierarchy problem}
We now consider the high energy physics above the Planck scale, especially asymptotically safe quantum gravity which is formulated as a nonperturbatively renormalizable quantum field theory~\cite{Hawking:1979ig,Reuter:1996cp,Souma:1999at}.
It is well-known that quantum gravity based on the Einstein-Hilbert action is not perturbatively renormalizable.
In other words, the Newton constant, which has canonical mass-dimension $-2$, is irrelevant at the Gaussian fixed point.
On the other hand, it is essential for the asymptotic safety scenario that gravitational couplings has a non-trivial UV fixed point at which the Newton constant is relevant.
A numerous work using the Wilsonian RG has shown evidences of the existence of such a fixed point.
See recent reviews~\cite{Eichhorn:2017egq,Percacci:2017fkn,Eichhorn:2018yfc,Reuter:2019byg,Wetterich:2019qzx}.

\begin{figure}[t]
\centering
\includegraphics[width=9cm]{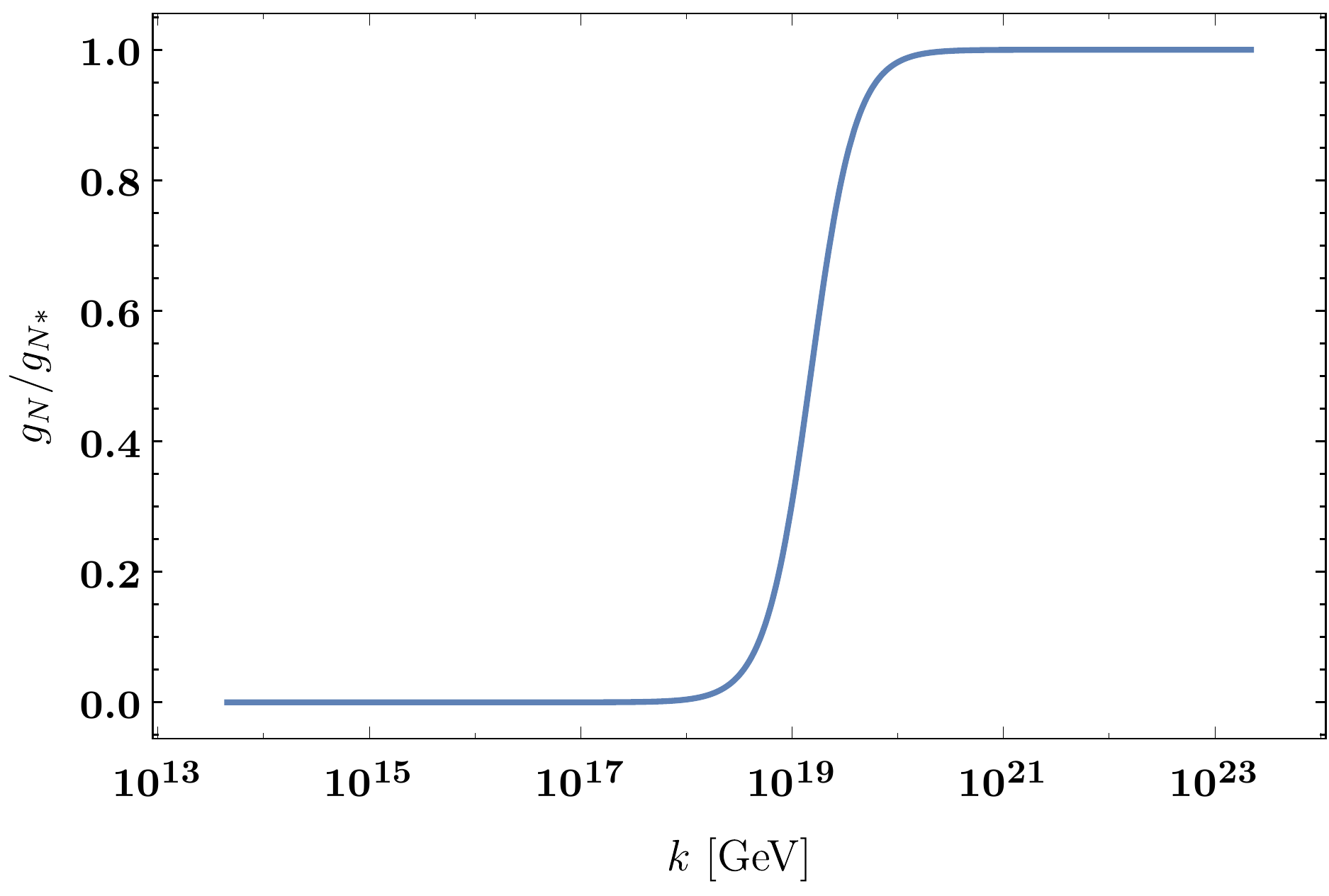}
\caption{Schematic figure for the RG flow of the dimensionless Newton constant.}
\label{RG flow of Newton constant} 
\end{figure}

We employ the Einstein-Hilbert truncation for the effective action,
\al{
\Gamma_k^\text{EH}=\int \df^4x \sqrt{g} \left[ \Lambda_\text{CC} -\frac{1}{16\pi G} R\right] +\Gamma_\text{gh} + \Gamma_\text{gf}\,,
}
where $\Lambda_\text{CC} $ is the cosmological constant; $G$ is the dimensionful Newton constant; $\sqrt{g}$ is the determinant of metric; $R$ is the Ricci scalar; and $\Gamma_\text{gh}$ and $\Gamma_\text{gf}$ are the ghost and gauge fixing actions for diffeomorphism, respectively.
The RG equation for the Newton constant is given by
\al{
\p_t g_N= (2 + \eta_g)g_N \,,
}
where $g_N=Gk^2$ is the dimensionless Newton constant.
Here, $\eta_g$ is the anomalous dimension induced by graviton fluctuations and has been computed. See e.g. Refs.\,\cite{Oda:2015sma,Hamada:2017rvn,Eichhorn:2017als,Pawlowski:2018ixd,Wetterich:2019zdo}.
The important fact is that $\eta_g$ takes a negative value smaller than $-2$ at the UV fixed point.
Therefore, the Newton constant becomes a relevant parameter at the UV fixed point, whereas we observe the irrelevant Newton constant with the critical exponent $\theta_{g_N}=-2$ ($\eta_g=0$) at the Gaussian fixed point, $g_{N*}=0$.
In Fig.\,\ref{RG flow of Newton constant}, we plot a schematic figure of the RG running of the dimensionless Newton constant with neglecting the cosmological constant.

One of advantages of asymptotically safe gravity is a strong predictability to the low energy dynamics of particles.  
Indeed, the Higgs mass of a mass of $125$\,GeV with a few GeV uncertainty was predicted before the discovery of the Higgs boson at the LHC~\cite{Shaposhnikov:2009pv}.
Besides, physical quantities such as quark masses could be predicted within the asymptotic safety program; see~\cite{Eichhorn:2017ylw,Eichhorn:2017muy,Eichhorn:2018whv,Eichhorn:2019dhg,Alkofer:2020vtb}.
It is crucial for these predictions how the anomalous dimensions induced by graviton fluctuations contribute to the beta functions of matter couplings.

Let us here investigate quantum gravity effects on the scalar potential. 
To this end, we replace the cosmological constant to the effective potential of a $O(N)$-scalar field $\phi$, i.e. $\Lambda_\text{CC} \to U(\phi)$, and add the kinetic term of the scalar field.
The RG equation for the effective potential taking into account only the quantum gravity effects reads~\cite{Pawlowski:2018ixd} 
\al{
\p_t \tilde U&=-4\tilde U + \tilde \phi \tilde U' 
+\frac{1}{24\pi^2}\left[ \frac{5}{1- v}+\frac{1}{1- v/4} -3\right]\,,
\label{beta function of effective scalar potential}
}
where we define dimensionless quantities, $\tilde \phi=\phi/k$, $\tilde U(\tilde \phi)= U(\phi)/k^4$ and $v=16\pi g_N \tilde U$, and the prime denotes the derivative with respect to $\tilde\phi$.
The first two terms on the right-hand side of Eq.\,\eqref{beta function of effective scalar potential} are the canonical scaling of the effective potential, while the third term is the graviton loop effect.
Expanding the effective potential into a polynomial of $\tilde\phi^2$, we obtain the RG equation for the scalar mass term,
 \al{
 \p_t {\bar m^2}&=\beta_m=(-2 +\gamma_m^g) {\bar m^2}\,,
\label{beta function of scalar mass}
 }
with the anomalous dimension induced by graviton fluctuations,
\al{
\gamma_m^g=\frac{g_N}{6\pi}\left[ \frac{20}{(1- v_0)^2}+\frac{1}{\left(1- v_0/4\right)^2}\right]\,,
}
where $v_0=16\pi g_N \tilde U(0)$ with $\tilde U(0)=\Lambda_\text{CC}/k^4$ the dimensionless cosmological constant.
From the RG equation \eqref{beta function of scalar mass} we find the critical exponent of the scalar mass term above the Planck scale,
\al{
\theta_m\simeq -\frac{\p \beta_m }{\p \bar m^2}\bigg|_{\substack{g_{N}=g_{N*} \\[0.5ex] v_0=v_{0*}}} = 2 - \frac{g_{N*}}{6\pi}\left[ \frac{20}{(1- v_{0*})^2}+\frac{1}{\left(1- v_{0*}/4\right)^2}\right]\,.
}
For a finite value of $g_{N*}$, the anomalous dimension becomes finite, so that the energy scaling of the scalar mass parameter is different from the canonical one.

\begin{figure}[t]
\centering
\includegraphics[width=9cm]{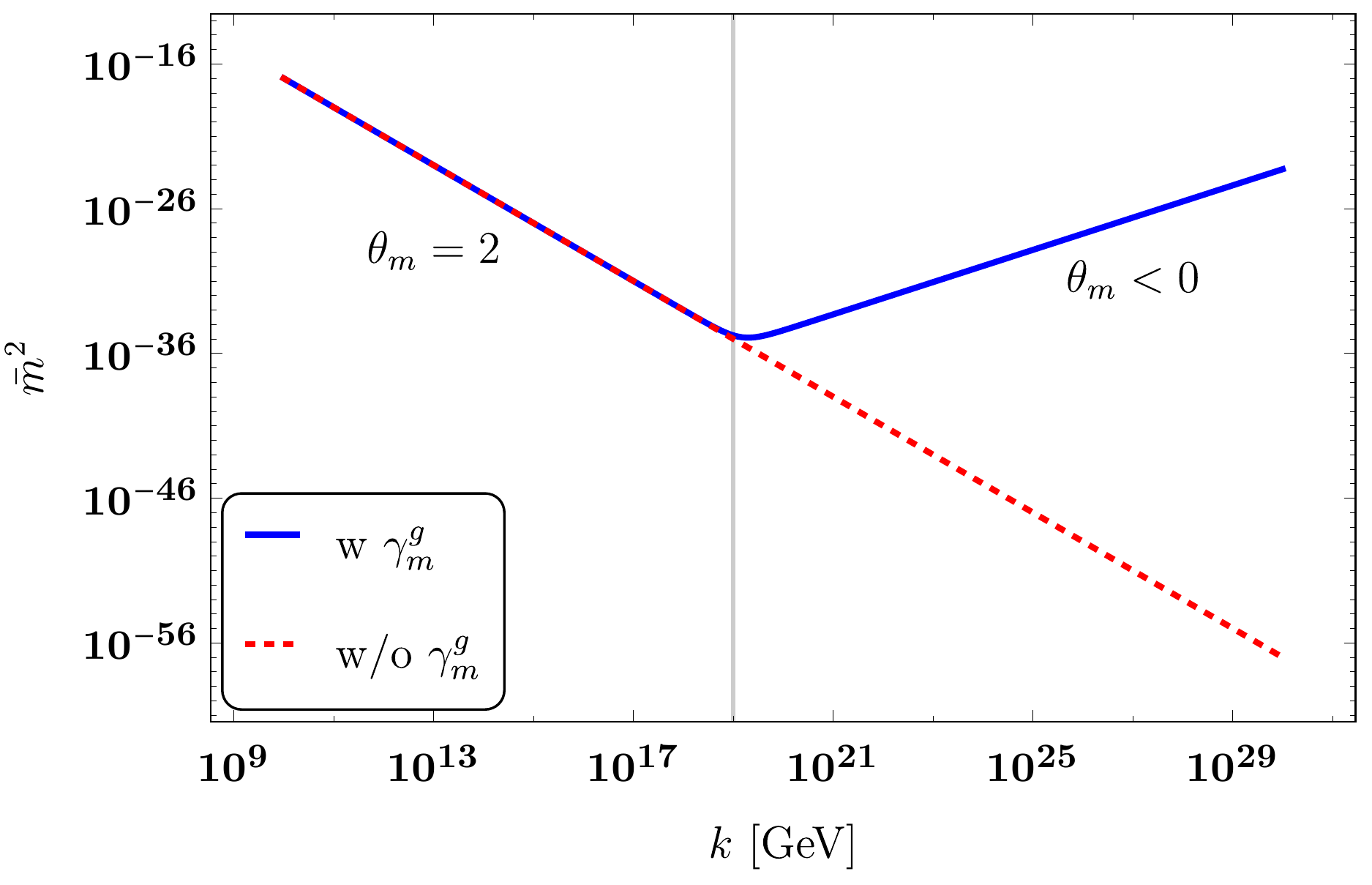}
\caption{Schematic figure for the RG flow of the scalar mass term with and without $\gamma_m^g>2$.}
\label{RG flow of scalar mass} 
\end{figure}

We here suppose that $\gamma_m^g>2$.
In this case, the sign of the critical exponent of the scalar mass becomes negative, namely the scalar mass parameter is irrelevant.This situation could provide possible solutions to the gauge hierarchy problem: One of them is the resurgence mechanism~\cite{Wetterich:2016uxm} in which the scalar mass parameter behaves as the blue solid lien in Fig.\,\ref{RG flow of scalar mass}. 
The scalar mass parameter shrinks towards zero in the regions above the Planck scale and then  increases such that $m_H^2/v_h^2\simeq 0.2$ at the electroweak scale due to the decoupling of quantum gravity effects below the Planck scale.
Hence, the Higgs mass parameter is tuned so as to be $m_0^2/M_\text{pl}^2\simeq 10^{36}$ around the Planck scale.
This is called as the self-tuned criticality~\cite{Bornholdt:1992up}.
The scalar mass parameter increases towards UV scales and then diverges. 
In order for the scalar mass (or equivalently the electroweak scale) to be UV safe and predictable (irrelevant), one has to have a quite small fixed point $\bar m^2_*\simeq 10^{-36}$.
So far, we have not found such a fixed point within simple truncated systems.
Instead, we find the Gaussian fixed point $\bar m^2_*=0$ for which the RG flow of the scalar mass keeps zero.
That is, the asymptotically safe condition in the continuum limit $k\to \infty$ implies that the low energy effective theory at the Planck scale could be (classically) scale invariant. 

We now give several comments in order.
The magnitude of the anomalous dimension $\gamma_m^g$ depends on the value of the fixed point of gravitational couplings, so that we need more precise analyze in order to establish the scenarios discussed above. 
In the scenarios with a negative critical exponent of the scalar mass parameter, the quadratic divergence is irrelevant: The UV theory is characterized by the UV fixed point and the low energy effective theory is given as a deviation from the fixed point.
That is, the quadratic divergence is subtracted.
However, the gauge hierarchy problem arises if there is a large intermediate scale between the Planck scale and the electroweak scale, e.g. the grand unification scale, $\Lambda_\text{GUT}$.
The large intermediate scale quadratically contributes to the Higgs mass parameter as discussed in Section~\ref{Sec: Scale from new physics}.

\section{Summary}
We have revisited the gauge hierarchy problem in elementary particle physics.
The quadratic divergence appearing in the scalar mass parameter may be spurious when we discuss the low energy dynamics of particles. 
The other aspect of the gauge hierarchy problem arises from the fact that the dimensionless scalar mass parameter $\bar m^2$ is a relevant parameter with the critical exponent $2$, namely behaves as $\bar m^2\sim k^{-2}$.
This fact with $m_H^2(v_h)/v_h^2\simeq 0.1$ implies that the Higgs mass at the Planck scale is quite smaller than the Planck scale: $m_H^2(M_\text{pl})/M_\text{pl}\simeq 10^{-36}$. 
A possible solution within the low energy physics is the imposition of classical scale invariance on the bare action.
Then, we extend the SM such that a scale $\Lambda_H$, which induces the electroweak scale, is generated.
In this paper, we have discussed such an extension of the SM based on the strongly interacting scalar-gauge theory in a hidden sector.
A newly added scalar field $S$ is coupled to the Higgs field $H$ via the Higgs-portal coupling $\lambda_{HS}(H^\dagger H)(S^\dagger S)$ which changes to the Higgs mass parameter proportional to the quadratic scale, i.e. $m_H^2\sim \lambda_{HS}\Lambda^2$.
This quadratic divergence is physical, namely independent from the choice of the regularization scheme.

Although classical scale invariance provides a direction of an extension of  the SM, it is a strong assumption.
To establish the scenario for electroweak scalegenesis, we have to discuss the high energy physics including quantum gravity beyond the Planck scale.
In this paper, we have introduced asymptotically safe quantum gravity which is realized as a non-perturbative field theory.
The anomalous dimensions induced by graviton fluctuations could drastically change the energy scaling of coupling constants above the Planck scale.
In particular, the graviton anomalous dimension $\gamma_m^g$ tends to make the critical exponent of the scalar mass parameter smaller than its canonical dimension.
If $\gamma_m^g$ is larger than 2, the scalar mass parameter becomes irrelevant.
In such a case, potential solutions to the gauge hierarchy problem are given: One of them is the resurgence mechanism in which the small Higgs mass parameter $\bar m_H^2/M_\text{pl}\simeq 10^{-36}$ is self-organized by quantum gravity effects.
The other is classical scale invariance.
That is, scale invariance at the Planck scale could be naturally realized as a consequence of the irrelevance of the scalar mass parameter above the Planck scale.
The gauge hierarchy problem may provide hints for a connection between low and high energy physics.

\subsection*{Acknowledgement}
The author thanks J.~Kubo and C.~Wetterich for collaborations and discussions.
He is supported by the Alexander von Humboldt foundation.


\begin{thebibliography}{99}
\bibitem{Gildener:1976ai} 
  E.~Gildener,
  Phys.\ Rev.\ D {\bf 14}, 1667 (1976).
\bibitem{Weinberg:1978ym} 
  S.~Weinberg,
  Phys.\ Lett.\  {\bf 82B}, 387 (1979).
  \bibitem{Aad:2012tfa}
  G.~Aad {\it et al.} [ATLAS Collaboration],
  Phys.\ Lett.\ B {\bf 716}, 1 (2012)
\bibitem{Chatrchyan:2012xdj} 
  S.~Chatrchyan {\it et al.} [CMS Collaboration],
  Phys.\ Lett.\ B {\bf 716}, 30 (2012)
    \bibitem{Wetterich:1983bi} 
  C.~Wetterich,
  Phys.\ Lett.\  {\bf 140B}, 215 (1984).
    \bibitem{Bardeen:1995kv} 
  W.~A.~Bardeen,
  FERMILAB-CONF-95-391-T.
  \bibitem{Coleman:1973jx} 
  S.~R.~Coleman and E.~J.~Weinberg,
  Phys.\ Rev.\ D {\bf 7}, 1888 (1973).
  doi:10.1103/PhysRevD.7.1888
\bibitem{Wetterich:1992yh} 
  C.~Wetterich,
  Phys.\ Lett.\ B {\bf 301}, 90 (1993)
  \bibitem{Berges:2000ew} 
  J.~Berges, N.~Tetradis and C.~Wetterich,
  Phys.\ Rept.\  {\bf 363}, 223 (2002)
  \bibitem{Litim:2001up} 
  D.~F.~Litim,
  Phys.\ Rev.\ D {\bf 64}, 105007 (2001)
  \bibitem{Gawedzki:1985cf} 
  K.~Gawedzki and A.~Kupiainen,
  Nucl.\ Phys.\ B {\bf 257}, 474 (1985).
  \bibitem{Aoki:2012xs} 
  H.~Aoki and S.~Iso,
  Phys.\ Rev.\ D {\bf 86}, 013001 (2012)
\bibitem{Sumi:2000xp}
  J.~I.~Sumi, W.~Souma, K.~I.~Aoki, H.~Terao and K.~Morikawa,
  hep-th/0002231.
\bibitem{Hamada:2012bp} 
  Y.~Hamada, H.~Kawai and K.~y.~Oda,
  Phys.\ Rev.\ D {\bf 87}, no. 5, 053009 (2013)
  Erratum: [Phys.\ Rev.\ D {\bf 89}, no. 5, 059901 (2014)]
  \bibitem{Martin:1997ns} 
  S.~P.~Martin,
  Adv.\ Ser.\ Direct.\ High Energy Phys.\  {\bf 21}, 1 (2010)
  [Adv.\ Ser.\ Direct.\ High Energy Phys.\  {\bf 18}, 1 (1998)]
  \bibitem{Veltman:1980mj} 
  M.~J.~G.~Veltman,
  Acta Phys.\ Polon.\ B {\bf 12}, 437 (1981).
  \bibitem{Wetterich:1981ir} 
  C.~Wetterich,
  Phys.\ Lett.\  {\bf 104B}, 269 (1981).
  \bibitem{Bornholdt:1992up} 
  S.~Bornholdt and C.~Wetterich,
  Phys.\ Lett.\ B {\bf 282}, 399 (1992).
  \bibitem{Wetterich:2011aa} 
  C.~Wetterich,
  Phys.\ Lett.\ B {\bf 718}, 573 (2012)
  \bibitem{Terao:2007jm} 
  H.~Terao and A.~Tsuchiya,
  arXiv:0704.3659 [hep-ph].
  \bibitem{Symanzik:1970rt} 
  K.~Symanzik,
  Commun.\ Math.\ Phys.\  {\bf 18}, 227 (1970).
  \bibitem{Callan:1970yg} 
  C.~G.~Callan, Jr.,
  Phys.\ Rev.\ D {\bf 2}, 1541 (1970).
  \bibitem{Lowenstein:1975rf} 
  J.~H.~Lowenstein and W.~Zimmermann,
  Commun.\ Math.\ Phys.\  {\bf 46}, 105 (1976).
  \bibitem{Haruna:2019zeu} 
  J.~Haruna and H.~Kawai,
  PTEP {\bf 2020},  no. 3,  033B01 (2020)
  \bibitem{Dermisek:2013pta} 
  D.~Chway, T.~H.~Jung, H.~D.~Kim and R.~Dermisek,
  Phys.\ Rev.\ Lett.\  {\bf 113}, no. 5, 051801 (2014)
  \bibitem{Kubo:2015cna} 
  J.~Kubo and M.~Yamada,
  Phys.\ Rev.\ D {\bf 93}, no. 7, 075016 (2016)
  \bibitem{Kubo:2014ova} 
  J.~Kubo, K.~S.~Lim and M.~Lindner,
  Phys.\ Rev.\ Lett.\  {\bf 113}, 091604 (2014)
  \bibitem{Kubo:2018vdw} 
  J.~Kubo and M.~Yamada,
  JHEP {\bf 1810}, 003 (2018)
    \bibitem{Kubo:2017wbv} 
  J.~Kubo, Q.~M.~B.~Soesanto and M.~Yamada,
  Eur.\ Phys.\ J.\ C {\bf 78}, no. 3, 218 (2018)
 \bibitem{Kubo:2015joa} 
  J.~Kubo and M.~Yamada,
  PTEP {\bf 2015}, no. 9, 093B01 (2015)
  \bibitem{Kubo:2016kpb} 
  J.~Kubo and M.~Yamada,
  JCAP {\bf 1612}, 001 (2016)
      \bibitem{Hawking:1979ig}
S. Weinberg, 
%
Chap. 16 in General Relativity ed. by Hawking, S.W. and Israel, W (1979).
    \bibitem{Reuter:1996cp} 
  M.~Reuter,
  Phys.\ Rev.\ D {\bf 57}, 971 (1998)
  \bibitem{Souma:1999at} 
  W.~Souma,
  Prog.\ Theor.\ Phys.\  {\bf 102}, 181 (1999)
    \bibitem{Eichhorn:2017egq} 
  A.~Eichhorn,
  Found.\ Phys.\  {\bf 48}, no. 10, 1407 (2018)
  \bibitem{Percacci:2017fkn} 
  R.~Percacci,
  ``An Introduction to Covariant Quantum Gravity and Asymptotic Safety,''
  \bibitem{Eichhorn:2018yfc} 
  A.~Eichhorn,
  Front.\ Astron.\ Space Sci.\  {\bf 5}, 47 (2019)
  \bibitem{Reuter:2019byg} 
  M.~Reuter and F.~Saueressig,
  ``Quantum Gravity and the Functional Renormalization Group : The Road towards Asymptotic Safety,''
  \bibitem{Wetterich:2019qzx} 
  C.~Wetterich,
  arXiv:1901.04741 [hep-th].
  \bibitem{Oda:2015sma} 
  K.~y.~Oda and M.~Yamada,
  Class.\ Quant.\ Grav.\  {\bf 33}, no. 12, 125011 (2016)
\bibitem{Hamada:2017rvn} 
  Y.~Hamada and M.~Yamada,
  JHEP {\bf 1708}, 070 (2017)
  \bibitem{Eichhorn:2017als} 
  A.~Eichhorn, Y.~Hamada, J.~Lumma and M.~Yamada,
  Phys.\ Rev.\ D {\bf 97}, no. 8, 086004 (2018)
  \bibitem{Pawlowski:2018ixd} 
  J.~M.~Pawlowski, M.~Reichert, C.~Wetterich and M.~Yamada,
  Phys.\ Rev.\ D {\bf 99}, no. 8, 086010 (2019)
  \bibitem{Wetterich:2019zdo} 
  C.~Wetterich and M.~Yamada,
  Phys.\ Rev.\ D {\bf 100}, no. 6, 066017 (2019)
  \bibitem{Shaposhnikov:2009pv} 
  M.~Shaposhnikov and C.~Wetterich,
  Phys.\ Lett.\ B {\bf 683}, 196 (2010)
  \bibitem{Eichhorn:2017ylw} 
  A.~Eichhorn and A.~Held,
  Phys.\ Lett.\ B {\bf 777}, 217 (2018)
  \bibitem{Eichhorn:2017muy} 
  A.~Eichhorn, A.~Held and C.~Wetterich,
  Phys.\ Lett.\ B {\bf 782}, 198 (2018)
  \bibitem{Eichhorn:2018whv} 
  A.~Eichhorn and A.~Held,
  Phys.\ Rev.\ Lett.\  {\bf 121}, no. 15, 151302 (2018)
  \bibitem{Eichhorn:2019dhg} 
  A.~Eichhorn, A.~Held and C.~Wetterich,
  arXiv:1909.07318 [hep-th].
  \bibitem{Alkofer:2020vtb} 
  R.~Alkofer, A.~Eichhorn, A.~Held, C.~M.~Nieto, R.~Percacci and M.~Schr{\" o}fl,
  arXiv:2003.08401 [hep-ph].
  \bibitem{Wetterich:2016uxm} 
  C.~Wetterich and M.~Yamada,
  Phys.\ Lett.\ B {\bf 770}, 268 (2017)
\end{thebibliography}

\end{document}